\documentclass[twocolumn]{aastex63}
\usepackage{amsmath}

\newcommand{\sphi}{\sin{\phi}}
\newcommand{\cphi}{\cos{\phi}}
\newcommand{\szeta}{\sin{\zeta}}
\newcommand{\czeta}{\cos{\zeta}}
\newcommand{\stheta}{\sin{\theta}}
\newcommand{\ctheta}{\cos{\theta}}
\newcommand{\eR}{{\boldsymbol e}_R}
\newcommand{\eRd}{{\boldsymbol e}_R^\prime}
\newcommand{\eO}{{\boldsymbol e}_\mathrm{O}}
\newcommand{\eS}{{\boldsymbol e}_\mathrm{S}}
\newcommand{\ThetaM}{\Theta_\mathrm{eq}}
\newcommand{\av}{{\boldsymbol a}}

\newcommand{\becv}{{\boldsymbol b}}
\newcommand{\xv}{{\boldsymbol x}}
\newcommand{\yv}{{\boldsymbol y}}
\newcommand{\lv}{{\boldsymbol l}}
\newcommand{\sv}{{\boldsymbol s}}
\newcommand{\pv}{{\boldsymbol p}}
\newcommand{\uv}{{\boldsymbol u}}
\usepackage{color}
\definecolor{red}{rgb}{0.0,0.0,0.0}
\usepackage{algorithmic}

\received{2020 March 1}
\revised{2020 April 3}
\accepted{2020 April 6}
\shorttitle{Global Composition Map of an Exo-Earth}
\shortauthors{Kawahara}
\graphicspath{{./}{figures/}}
\begin{document}

\title{Global Mapping of the Surface Composition on an Exo-Earth using Color Variability}
 \correspondingauthor{Hajime Kawahara}
 
\author[0000-0003-3309-9134]{Hajime Kawahara}
\email{kawahara@eps.s.u-tokyo.ac.jp}
\affiliation{Department of Earth and Planetary Science, The University of Tokyo, 7-3-1, Hongo, Tokyo, Japan}
\affiliation{Research Center for the Early Universe, 
School of Science, The University of Tokyo, Tokyo 113-0033, Japan}
\nocollaboration{1}

\begin{abstract}
Photometric variation of a directly imaged planet contains information on both the geography and spectra of the planetary surface. We propose a novel technique that disentangles the spatial and spectral information from the multi-band reflected light curve. This will enable us to compose a two-dimensional map of the surface composition of a planet with no prior assumption on the individual spectra, except for the number of independent surface components. We solve the unified inverse problem of the spin-orbit tomography and spectral unmixing by generalizing the non-negative matrix factorization (NMF) using a simplex volume minimization method. We evaluated our method on a toy cloudless Earth and observed that the new method could accurately retrieve the geography and unmix spectral components. Furthermore, our method is also applied to the real-color variability of the Earth as observed by Deep Space Climate Observatory (DSCOVR). The retrieved map explicitly depicts the actual geography of the Earth and unmixed spectra capture features of the ocean, continents, and clouds. It should be noted that, the two unmixed spectra consisting of the reproduced continents resemble those of soil and vegetation. 
\end{abstract}

\keywords{astrobiology -- Earth -- scattering -- techniques: photometric, nonnegative matrix factorization}

\section{Introduction}

Direct imaging of terrestrial planets around a nearby solar-type star are important targets in future astronomy. In the 2020 decadal surveys, both {\it HabEx} and {\it LUVOIR} have shown a capability to search for these planets, even in the habitable zone. Direct imaging with spectroscopy provides information regarding the molecules in the atmosphere of the planet, which enables us to search for biosignatures such as oxygen, carbon dioxide, and water. Moreover, surface inhomogeneity can be explored with photometric monitoring of the reflected light as proposed by \cite{2001Natur.412..885F}. The color variability of the reflected light has been studied as a probe of surface compositions \citep[e.g.][]{2009ApJ...700..915C,2011ApJ...738..184F}. The spatial distribution of the planet surface can also be inferred from the photometric variation. Diurnal variation, due to the rotation of the planet, provides the spin rotation period and a one-dimensional distribution of the surface \citep{2008ApJ...676.1319P,2009ApJ...700..915C, 2009ApJ...700.1428O, 2010ApJ...715..866F, 2011ApJ...738..184F,2018AJ....156..301L}. Furthermore, the axial tilt can be obtained from the analysis of the frequency modulation of the periodicity \citep{2016ApJ...822..112K, Nakagawa}. The analytic expression of the reflected light curves has been studied \citep{2013MNRAS.434.2465C,2018MNRAS.478..371H}.

A full two-dimensional inversion technique called ``spin-orbit tomography'' (analogous to computer tomography), was proposed by \cite{2010ApJ...720.1333K}, and has been studied in terms of the inverse problem \citep{2011ApJ...739L..62K,2012ApJ...755..101F,2018AJ....156..146F, 2019AJ....158..246B, Aizawa} and obliquity measurement \citep{2016MNRAS.457..926S,2018AJ....156..146F}. Recently, \cite{2019arXiv190312182L} analyzed a single-band light curve of the Earth with Transiting Exoplanet Survey Satellite (TESS) data and inferred a rough two-dimensional cloud distribution. \cite{2019ApJ...882L...1F} successfully retrieved a global map that was analogous to the distribution of a continent, from data that was obtained by DSCOVR by monitoring the Earth for two years \citep{2018AJ....156...26J}. They used the second principle component (PC2) of a multi-color light curve. \cite{Aizawa} improved the retrieved map from DSCOVR using sparse modeling. These examples showed that a global map could be retrieved from a time-series of a single band or PC. However, there exists a level of ambiguity when interpreting the derived maps when we do not have prior knowledge on the surface compositions.

Moreover, a blind retrieval of the reflectance spectra of the surface components from the integrated light is known as ``spectral unmixing'' in remote sensing. \cite{2013ApJ...765L..17C} formulated the spectral unmixing as a disentanglement of geography by spin rotation. However, the longitudinal map inferred from the EPOXI data did not match with the real geographies because of the degeneracy of the inferred geometric distribution and spectral components \citep{2017AJ....154..189F}. The ambiguity of spectral unmixing originates from the matrix factorization not being unique, which has been extensively studied in the field of remote sensing. These studies found that additional constraints such as the simplex volume minimization of spectral components guarantee a unique solution to the unmixed spectra \citep{1994ITGRS..32..542C,2015ITSP...63.2306F,2015ITGRS..53.5530L,2019ISPM...36...59F,2019arXiv190304362A}. In practice, non-negative matrix factorization (NMF) with regularization terms easily retrieve the surface components in hyperspectral unmixing \citep{2019arXiv190304362A}. These techniques in remote sensing are worth considering in their application to multi-color light curves of directly imaged exoplanets.

This paper aims to formulate a single inverse problem that unifies the spin-orbit tomography and spectral unmixing using a novel technique used in remote sensing. To achieve this, we unify the NMF-based spectral unmixing technique and spin-orbit tomography to retrieve both the spectra and geographies of a disk-integrated light curve from an exoplanet. We demonstrate its capabilities using the simulated data and real data from Deep Space Climate Observatory (DSCOVR). The rest of the paper is organized as follows. In Section 2,  we first review the spin-orbit tomography and spectral unmixing. Next, we construct a unified retrieval model using NMF; the optimization scheme is also provided. In Section 3, we test the technique by applying it to a cloudless toy model. In Section 4, we demonstrate this new technique by applying it to real observational data of the Earth recorded by DSCOVR. Finally, in Section 5, we summarize our results.

\section{Formulation of Spin-Orbit Tomography with Spectral Unmixing}

\subsection{Spin-Orbit Tomography}
Space direct imaging in optical and near-infrared bands aim to detect reflected lights (or scattered lights) of a host star near a planet. The reflected light is a summation of photons from a day and visible side of a planet. This integrated-reflected light is expressed as   
\begin{eqnarray}
\label{eq:brdfdef4xX}
\label{eq:integrate}
f_p = \frac{f_\star R_p^2}{\pi a^2} \int_{\mathrm{IV}} d \Omega_1  R^s(\vartheta_0,\vartheta_1,\varphi)  \cos{\vartheta_0} \cos{\vartheta_1},
\end{eqnarray}
where $f_\star$ is the stellar flux, $R_p$ is the radius of the planet, $a$ is the star-planet distance, IV is the illuminated and visible area, and $\Omega_1$ is the solid angle of the planet’s sphere. $ R^s(\vartheta_0,\vartheta_1,\varphi)$ represents the bidirectional reflectance distribution function (BRDF) of the surface element $s$. $\vartheta_0$ is the solar zenith angle, ${\vartheta_1}$ is the zenith angle between the direction towards an observer and  normal vector of the surface, and $\varphi$ is the relative azimuth angle between the line-of-sight and stellar direction. The derivation of equation (\ref{eq:integrate}) is given in Appendix \ref{ap:sot}. An isotropic approximation of the surface reflectance (the Lambert approximation), $ R^s(\vartheta_0,\vartheta_1,\varphi) = R^s$, significantly reduces the complexity of the problem. We also define the spherical coordinate fixed on the surface by $(\theta, \phi)$ and express the surface component $s$, in spherical coordinates $R^s = m(\theta,\phi)$ as the time-independent quantity (static surface approximation). Then, we obtain
\begin{eqnarray}
\label{eq:brdfdef4xX}
f_p = \frac{f_\star R_p^2}{\pi a^2} \int_{\mathrm{IV}} d \Omega_1  m(\theta, \phi)  \cos{\vartheta_0} \cos{\vartheta_1}.
\end{eqnarray}
We note that the IV area, $\cos{\vartheta_0}$, and $ \cos{\vartheta_1}$ are time-dependent. The terms of $\cos{\vartheta_0}$ and $ \cos{\vartheta_1}$ also depend on the position of the planet surface, $(\theta, \phi)$, and the axial tilt parameters, ${\bf g} = (\zeta, \ThetaM)$, {\color{red} where $\zeta$ is the planet obliquity and $\ThetaM$ is the orbital phase at the equinox.} We define the geometric kernel introduced by \cite{2010ApJ...720.1333K} as  
\begin{eqnarray}
\label{eq:brdfdef4xX}
W_{\bf g}(t,\theta,\phi) = 
  \left\{
    \begin{array}{l}
\displaystyle{\frac{f_\star R_p^2}{\pi a^2}  \cos{\vartheta_0} \cos{\vartheta_1} \mbox{ for $\cos{\vartheta_0}, \cos{\vartheta_1}>0$}}\\
\\
\displaystyle{ 0 \mbox{\,\, otherwise,} }
    \end{array}
  \right.
\end{eqnarray}
where the positive condition of $\cos{\vartheta_0}$ and $\cos{\vartheta_1}$ restrict the surface integral to pixels on the IV area. Assuming that {\bf g} is fixed, we obtain the Fredholm integral equation of the first kind
\begin{eqnarray}
\label{eq:fredfirst}
f_p (t) = \int d \Omega  \, W_{\bf g}(t,\theta,\phi) \, m(\theta, \phi).
\end{eqnarray}

Discretization of the time $t \to t_i $ and planetary surface $(\theta, \phi) \to (\theta_j, \phi_j) $ reduces the equation (\ref{eq:fredfirst}) to the linear inverse problem
\begin{eqnarray}
\label{eq:sotij}
d_i = \sum_j  W_{ij} m_{j},
\end{eqnarray}
or using the vector form, we can express it as
\begin{eqnarray}
\label{eq:sot}
{\bf d} = W {\bf m},
\end{eqnarray}
where $d_i = f_p (t_i)$  for $i=0,1,...,N_i-1$ and $m_j = m(\theta_j, \phi_j)$ for $j=0,1,...,N_j-1$. The explicit expression of the geometric kernel $W_{ij} =  W_{\bf g}(t_i,\theta_j,\phi_j)$ in the spherical coordinate is given in Appendix \ref{ap:sot}.

Because the inverse problem (\ref{eq:sot}) is ill-posed, an additional constraint or regularization is needed to solve the problem. Various types of regularizations have been attempted so far. \cite{2010ApJ...720.1333K} used non-negative regularization, and the requirement of an upper limit of albedo as regularization using the bounded variable least squares solver \citep{lawson1995solving}. \cite{2011ApJ...739L..62K} used the Tikhonov regularization, which minimizes the cost function  
\begin{eqnarray}
\label{eq:sotcost}
\hbox{minimize \,} Q = \frac{1}{2}||{\bf d} - W {\bf m}||_2^2 + \frac{\lambda_A}{2} ||{\bf m}||_2^2,
\end{eqnarray}
where $\lambda_A$ is the spatial regularization parameter and $||\cdot||_2^2$ is the squared L2 norm. To construct the model on the Bayesian framework, \cite{2018AJ....156..146F} used a Gaussian process to regularize the map while \cite{2019AJ....158..246B} used an Occamian approach algorithm. Recently, \cite{Aizawa} demonstrated that the L1 + total square variation (TSV) provided better results than a simple L2 (Tikhonov) regularization. 

The value of ${\bf d}$ depends on what features we want to extract from the multi-color light curve. \cite{2011ApJ...739L..62K} used a single-band light curve to retrieve a cloud map of the simulated Earth. They also demonstrated that a rough two-dimensional distribution of the continent or ocean can be retrieved from a color difference between 0.85 micron and 0.45 micron or 0.85 micron and 0.65 micron, owing to the near flatness of the cloud spectrum. {\color{red} \cite{2009ApJ...700..915C} utilized principle component analysis (PCA) for their longitudinal mapping of EPOXI data.} \cite{2019ApJ...882L...1F} used the second component of PCA of the multi-color light curve of DSCOVR. Comparing with the ground truth, they found that the resultant map was similar to the global continent/ocean map of the Earth. However, these two examples required prior knowledge of the surface composition or the ground truth of the geography. The ambiguity in the interpretation of the map is a limitation of the spin-orbit tomography.

\subsection{Spectral Unmixing}\label{ss:sumix}

Spectral unmixing is a procedure that disentangles mixed spectra by finding the endmembers. The mixing model of the spectra of multiple surface compositions is required to unmix the spectra. The simplest model is the linear mixing model, expressed as 
\begin{eqnarray}
\label{eq:suij}
d(t_i, \tilde{\lambda}_l) = D_{il} = \sum_k  A_{ik} X_{kl},
\end{eqnarray}
or simply
\begin{eqnarray}
\label{eq:su}
D = A X,
\end{eqnarray}
where $A_{ik} = a_k(t_i)$ is the contribution of the $k$-th component at time $t=t_i$ to the intensity of light and $X_{kl} = x_k(\tilde{\lambda}_l)$ for $l=0,1,...,N_l-1$ is the reflection spectra of the $k$-th component at wavelength $\tilde{\lambda}_l$\footnote{We note that the spectral unmixing in remote sensing is often expressed in the form of $D^\prime = X^\prime (A^\prime)^T$ instead of equation (\ref{eq:su}) , i.e. ``spectral component first'', where $D^\prime = D^T, X^\prime = X^T, A^\prime = A$. We select the form of equation (\ref{eq:su}) because of the connectivity between the unmixing and spin-orbit tomography as seen in section \ref{ss:uni}.}. We need to solve the matrix factorization of $A$ and $X$. Generally, the matrix factorization is formulated as the minimization of the cost function, where the cost function can either be the squared Euclidean distance or the Kullback--Leibler distance. In this paper, we use the squared Euclidean distance
\begin{eqnarray}
\label{eq:cost_matfact}
 Q = \frac{1}{2}|| D - A X ||_F^2 + R(A, X)
\end{eqnarray}
where $|| \cdot ||_F^2$ is the squared Frobenius norm defined by 
\begin{eqnarray}
|| Y ||_F^2 \equiv \sum_j \sum_i Y_{ij}^2.
\end{eqnarray}
and $R(A, X)$ is the regularization term.

\subsubsection{Principle Component Analysis}
Principle component analysis (PCA) is a traditional technique used to disentangle the spectral components of multi-color light curves as observed in \cite{2009ApJ...700..915C}. It was also used in a global map reconstruction of the Earth by \cite{2019ApJ...882L...1F} and \cite{Aizawa}. PCA can also be formulated as a minimization of the cost function, from the perspective of optimization, 
\begin{eqnarray}
\label{eq:PCA}
\hbox{minimize \,} Q &=& \frac{1}{2}|| D - A X ||_F^2 \\
\hbox{ \,\, subject to } 
A^T A &=& \mathrm{diag}{(\bf \sigma_A)} = \Sigma_A,\\
X^T X &=& \mathrm{diag}{(\bf \sigma_X)} = \Sigma_X,
\end{eqnarray}
where $\mathrm{diag}{(\bf \sigma)}$ is a diagonal matrix whose elements are $\sigma_i$.   The drawback of the PCA as a matrix factorization method is the strong assumption of orthogonality for $A$ and $X$. However, its orthogonality is useful to visualize the simplex by reducing its dimensionality {\color{red} \citep{2013ApJ...765L..17C}}. In this paper, we denote the orthogonal PCA basis by $U_X = (\Sigma_X^{-1/2} X)^T$, i.e. $U_X^T U_X = I$ ($I$ is an identity matrix). An arbitrary matrix $M$ can be decomposed by {\color{red} row} vectors of $U_X$ as
\begin{eqnarray}
\label{eq:PCAdec}
M &=& \sum_k \pv_{k} \uv_k^T,
\end{eqnarray}
where $\uv_k$ is the $k$-th {\color{red} row} of $U_X$. The projection of $M$ on to PC$k$ is computed by 
\begin{eqnarray}
\label{eq:PCAproj}
\pv_{k} &=& M \uv_k.
\end{eqnarray}

\subsubsection{Non-negative Matrix Factorization}\label{ss:nmf}

In the field of remote sensing, a wide variety of spectral unmixing has been studied. Among these techniques, NMF decomposes a single matrix $D$ to two matrices $A$ and $X$ whose elements are non-negative, that is, $D = A X$ \citep{paatero1994positive,lee2001algorithms}. NMF can be defined by the minimization of the cost function. 
For instance, using the squared Euclidean distance, NMF is formulated as
\begin{eqnarray}
\label{eq:NMF}
\hbox{minimize \,} Q = \frac{1}{2}|| D - A X ||_F^2 + R(A, X)\\
\hbox{ \,\, subject to } A_{ik} \ge 0, X_{kl} \ge 0.
\end{eqnarray}
NMF is known to be NP-hard \citep{vavasis2009complexity}; therefore, the optimization of NMF is difficult to achieve. Nevertheless, various efficient optimization methods have been proposed \citep[][references therein]{lee2001algorithms, cichocki2009nonnegative}. 

%
%

\begin{figure}[]
 \begin{center}
   \includegraphics[width=\linewidth]{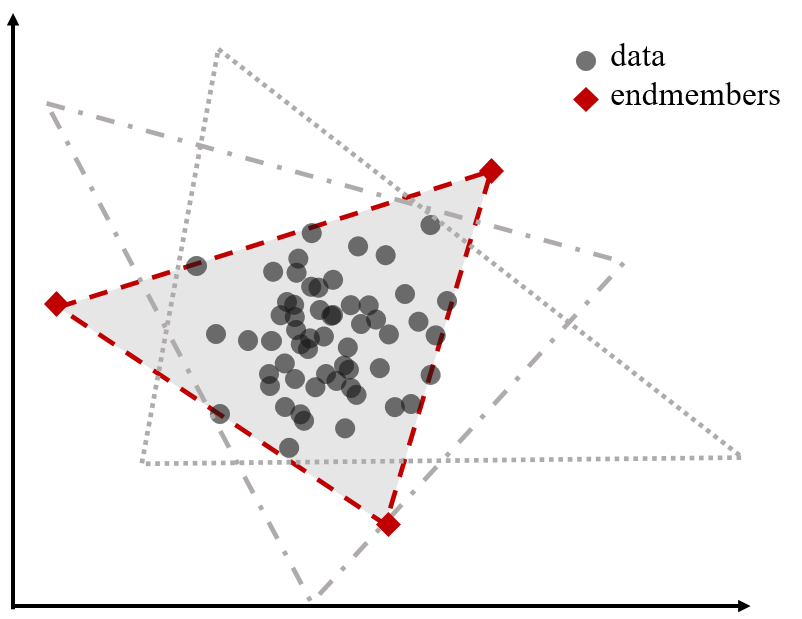}
 \end{center}
 \caption{Schematic picture of the simplex volume minimization, {\color{red} which is based on Figure 1 in \cite{2015ITGRS..53.5530L} \citep[see also][]{2017AJ....154..189F}}. The black dots represent the observed data and the three triangles indicate a simplex that encloses all of the data points. The dashed triangle is the simplex whose volume is minimized. The end members are defined by three vertices of the dashed triangle.  \label{fig:volmin}}
\end{figure}

In particular, NMF combined with the simplex volume minimization technique can accurately reproduce the high-resolution spectrum components from remote-sensing satellite data \citep{1994ITGRS..32..542C,2015ITSP...63.2306F,2015ITGRS..53.5530L,2019ISPM...36...59F,2019arXiv190304362A}. The concept of the simplex volume minimization can be summarized as follows: If the data are sufficiently spread in the convex hull defined by the end members, the data-enclosing simplex whose volume is minimized identifies the true end members. 

In Figure \ref{fig:volmin}, we plot three simplexes that enclose all of the data points. Each simplex provides its vertices as a solution of NMF. The simplex volume minimization choose the vertices of the volume-minimum simplex (dashed triangle) as the end members of NMF. {\color{red} When there is at least one pure pixel of each end member in the data, the volume-minimum simplex obviously identifies the true end members. Even in the case without pure pixels, \cite{2015ITGRS..53.5530L} showed that the true end members could be identified by the volume-minimum simplex under the condition of the pixel purity level that applies uniformly to all of the end members. In Figure \ref{fig:volmin}, if the data points on the red dashed lines have high purity levels, that is, they are on the boundaries of the simplex defined by the true endmembers, then the volume-minimum simplex identifies the true endmembers. 
}

As the regularization term for the simplex volume minimization, the Gram determinant of spectral components (VRDet)
\begin{eqnarray}
\label{eq:vrterm}
R(A,X) &=& \frac{\lambda_X}{2} \, \det {(X X^T)} \\
&=& \frac{\lambda_X}{2} \, \det_{k,k^\prime} \left[ { \sum_l (X_{kl} X_{k^\prime l})} \right],
\end{eqnarray}
was used \citep[e.g.][]{schachtner2009minimum,zhou2011minimum,xiang2015blind,2019ISPM...36...59F,2019arXiv190304362A}, where $\lambda_X$ is the spectral regularization parameter. The Gram determinant (\ref{eq:vrterm}) is a surrogate of the volume of a convex hull of spectral vectors, $( {\bf x}_0, {\bf x}_1,  ....  {\bf x}_{N_k-1} )$, where ${\bf x}_k = \{ x_{k} (\tilde{\lambda}_l)$ for $l=0,1,...,N_l-1 \}$; such a convex hull is identifiable for its well-spread data \citep{2015ITGRS..53.5530L}\footnote{\color{red} The Gram determinant can be rewritten by the wedge product of the spectral vectors as $\det {(X X^T)} = ||\xv_0 \wedge \xv_1 \wedge ... \wedge \xv_{k-1}||^2$. Therefore, $\det {(X X^T)}$ can be regarded as the squared volume of the spectral vectors. }. The minimization of the convex hull of spectral vectors can be achieved by minimizing (\ref{eq:vrterm}).

\subsection{Unified Retrieval Method of Mapping and Spectra}\label{ss:uni}

Our task is to unify the spectral unmixing of Equation (\ref{eq:su}) and spin-orbit tomography of Equation (\ref{eq:sot}). To achieve this, we assume a pixel-wise spectral unmixing
\begin{eqnarray}
\label{eq:unmix}
m(\theta_j, \phi_j, \tilde{\lambda}_l) = m_{jl} &=& \sum_k  A_{jk} X_{kl},
\end{eqnarray}
where $X_{kl}$ is the reflectivity of the $k$-th component at wavelength $\tilde{\lambda}_l$, and $A_{jk} = a_k (\theta_j, \phi_j)$ is the surface distribution of the $k$-th surface component at the $j$-th pixel instead of time in equation (\ref{eq:su}). Combining Equation (\ref{eq:unmix}) with the multicolor version of equation (\ref{eq:sot}), we obtain 
\begin{eqnarray}
D_{il} = \sum_j W_{ij} m_{jl} =  \sum_{jk} W_{ij} A_{jk} X_{kl}
\end{eqnarray}
or simply,
\begin{eqnarray}
\label{eq:general_sotsu}
D = W A X.
\end{eqnarray}
Equation (\ref{eq:general_sotsu}) provides the general form of the spin-orbit tomography with spectral unmixing. 

The two-dimensional mapping thus far estimated the spectra using PCA or a color difference prior to retrieving the geographic distribution. This ``unmixing first'' strategy could not feed back information on the fitting accuracy of the geographic retrieval to spectral unmixing. The improvement of Equation (\ref{eq:general_sotsu}) over the spin-orbit tomography is that we fit both the spectral components and geography to data in a consistent manner.

\cite{2013ApJ...765L..17C} solved an equation similar to (\ref{eq:general_sotsu}) for $A$ and $X$ using the multicolor light curve provided by the EPOXI satellite as $D$. They retrieved a longitudinal map of the surface components from the diurnal rotation of the light curve. This procedure is referred to as ``rotational unmixing''. They used the Markov Chain Monte Carlo method to determine the best parameters of $X$ and $D$. Their optimization corresponds to the minimization of 
\begin{eqnarray}
\label{eq:general_sotsu_mini}
&\,&  Q = \frac{1}{2}|| D - \overline{W} A X ||_F^2 \\
\label{eq:general_sotsu_mini_c}
&\,& \hbox{ \,\, subject to }  1 \ge  \sum_k A_{jk} \ge 0, 
1 \ge X_{kl} \ge 0,
\end{eqnarray}
where $\overline{W}$ is the latitudinal average of the kernel\footnote{Besides the additional constraints on $X$ and $A$, the difference between the equation in \cite{2013ApJ...765L..17C} and equation (\ref{eq:general_sotsu}) is the $W$. As rotational unmixing performs the longitudinal mapping according to spin rotation, the geometric kernel should be integrated unto the latitudinal direction, $\overline{W} = \overline{W}(\phi)$. In the frame of the spin-orbit tomography, we need to use a two-dimensional discretization of a sphere, $W(\theta, \phi)$.}. The minimization of equation (\ref{eq:general_sotsu_mini}) under the constraint of Equation (\ref{eq:general_sotsu_mini_c}) is formally identical to the weighted NMF (we explain this in \S \ref{ss:WNMF}) with {\it no} regularization + the upper limits of $A$ and $X$.

In general, matrix factorization has a degeneracy of solutions. The transformation of $A \leftarrow A G^{-1} $, and $X \leftarrow G X$ for a regular matrix $G$ under the constraint of Equation  (\ref{eq:general_sotsu_mini_c}) does not change the value of the cost function of Equation (\ref{eq:general_sotsu_mini}). This means that if we change the spectral basis by a rotation of $G$, then the inferred map should change too. This degeneracy is known in the field of blind signal separation \citep[e.g. see Chapter 1.3.2 in ][]{cichocki2009nonnegative}. 
The nonuniqueness of the blind signal separation might explain the mismatch between the inferred longitudinal map by \cite{2013ApJ...765L..17C} and the actual geography suggested by \cite{2017AJ....154..189F}. 

The nonuniqueness feature of NMF can be avoided by adding regularization when neglecting unavoidable scaling and permutation ambiguities \citep{cichocki2009nonnegative,2015ITGRS..53.5530L}. Our task is to find a unique (identifiable) solution to the spectral unmixing and spin-orbit tomography. To achieve this, we consider the cost function with regularization for both $X$ and $A$. In this paper, we used the squared Euclidean distance as the cost function for equation (\ref{eq:general_sotsu})
\begin{eqnarray}
\label{eq:cost}
 Q = \frac{1}{2}|| D - W A X ||_F^2 + R(A,X)
\end{eqnarray}
where $R(A,X)$ is the regularization term. Similar to the rotation unmixing for longitudinal mapping \citep{2013ApJ...765L..17C}, we call the two-dimensional mapping+unmixing+regularization of Equation (\ref{eq:cost}) “spin-orbit unmixing” in this paper. 

\subsection{Weighted Nonnegative Matrix Factorization}\label{ss:WNMF}
The nonnegative condition of Equation (\ref{eq:cost}) yields an NMF version of the unified retrieval model  
\begin{eqnarray}
\label{eq:WAX}
\hbox{minimize \,} Q = \frac{1}{2}|| D - W A X ||_F^2 + R(A,X) \\
\label{eq:WAXc}
\hbox{ \,\, subject to } A_{jk} \ge 0, X_{kl} \ge 0.
\end{eqnarray}
As this formulation differs from a standard NMF, Equation (\ref{eq:NMF}) for a  weight $W$, we require an extension for the optimization of a standard NMF to the weighted NMF. 

\subsubsection{Regularization}

The regularization term suppresses the instability of the retrieved map due to overfitting, otherwise known as over-training in machine learning. A Tikhonov regularization (or a L2 regularization) used in the original spin-orbit tomography \citep{2011ApJ...739L..62K} can be extended to the regularization term using the Frobenius norm, $R(A,X) \propto ||A||_F^2$. Hence, a simple extension of the spin-orbit tomography is expressed as
\begin{eqnarray}
\label{eq:vrL2x}
R(A,X) = \frac{\lambda_A}{2} ||A||_F^2 \hbox{\,\, (L2-{Unconstrained})}.
\end{eqnarray}
However, in the case of ``L2'', we do not have any regularization for $X$ (``L2-Unconstrained''). In spectral unmixing, an assumption made on the convex hull of spectral components can be expressed as a regularization term (a function of $X$), as explained in Section \ref{ss:nmf}.  We consider a combination of the Tikhonov regularization for mapping, and the Gram determinant-type volume regularization for spectral components, expressed as
\begin{eqnarray}
\label{eq:vrnmfreg}
R(A,X) = \frac{\lambda_A}{2} ||A||_F^2 + \frac{\lambda_X}{2} \det{(X X^T)} \nonumber \\
\hbox{\,\, (L2-VRDet)},
\end{eqnarray}
where  $\lambda_A$ and $\lambda_X$ the are the regularization parameters for $A$ and $X$, respectively. We call this model ``L2-VRDet''.

\subsubsection{Optimization}

The minimization of the cost function is performed by a block coordinate descent, which consists of two separate optimizations for $A$ and $X$ \citep[see][as a review paper]{kim2014algorithms}. These optimizations are solved by minimizing the quadratic forms
\begin{eqnarray}
q_A &=& \frac{1}{2} \av^T_k \mathcal{W}_A \av_k - \becv_A^T \av_k \\
q_X &=& \frac{1}{2} \xv^T_k \mathcal{W}_X \xv_k - \becv_X^T \xv_k.
\end{eqnarray}
 of $\av_k$ (the $k$-th {\color{red} column} vector of $A$) and $\xv_k$ (the $k$-th {\color{red} row} vector of $X$) for $k=0$ to $N_k-1$, iteratively \citep{zhou2011minimum}{\color{red}, where $\mathcal{W}_A$,$\mathcal{W}_X$,$\becv_A$, and $\becv_X$ are placeholders that depend on the cost function. }
 
As the L2 regularization of $A$, we minimize 
\begin{eqnarray}
q_A &=& \frac{1}{2} \av^T_k (\mathcal{L}_A + \mathcal{T}_A)\av_k - \lv_A^T \av_k \,\, \mbox{(L2)}\\
 \mathcal{L}_A &=& \xv_k^T \xv_k  W^T W \\
 \lv_A &=& W^T \Delta \, \xv_k \\
 \mathcal{T}_A &=& \lambda_A I_J
\end{eqnarray}
where $\Delta_{il} =  D_{il} - \sum_{s \neq k} \sum_{j} W_{ij} A_{js} X_{sl}$, $I_J \in \mathbb{R}^{N_j \times N_j}$ is an identity matrix . The $X$ component of the quadratic problem for the VRDet model is calculated by  
\begin{eqnarray}
q_X &=& 
\displaystyle{\,\frac{1}{2} \xv^T_k (\mathcal{L}_X + \mathcal{D}_X) \xv_k - \lv_X^T \xv_k}  \mbox{\,\,(VRDet),} \\
\mathcal{L}_X &=& ||W \av_k||_2^2 \, I_L \\
\lv_X &=& \Delta^T W \av_k \\
\mathcal{D}_X &=& \lambda_X \det{(\breve{X}_k \breve{X}_k^T)} [ I_L - \breve{X}_k^T  (\breve{X}_k \breve{X}_k^T)^{-1} \breve{X}_k ],
\end{eqnarray}
where $\breve{X}_k$ is a submatrix of $X$ when the $k$-th row of $X$ is removed, and  $I_L \in \mathbb{R}^{N_l \times N_l}$ is the identity matrix. The derivations of these terms are given in Appendix \ref{ap:qp}.

Following \cite{2019arXiv190304362A}, we use the accelerated projected gradient descent + restart (APG+restart) to optimize the quadratic problems with nonnegative conditions. The APG+restart is based on a projected gradient descent onto a positive orthant with Nesterov's acceleration and the restarting method. The algorithm is summarized as follows. 

\begin{algorithmic}
\STATE 
\STATE {\bf Algorithm:} \underline{NMF/Block Coordinate Descent for}
\STATE \underline{Spin-Orbit Unmixing}
\STATE
\STATE Minimize $\frac{1}{2}||D-WAX||_F^2 + R(A,X)$ s.t. $A,X \ge 0$
\STATE Initialize $A^{(0)},X^{(0)}$  by random nonnegative values
\WHILE {Condition} 
    \FOR {k in ($0,N_k-1$)} 
        \STATE Update $\xv_k$ using APG+restart
        \STATE Update $\av_k$ using APG+restart
    \ENDFOR
\ENDWHILE
\end{algorithmic}
\vspace{\baselineskip}
A more detailed description of the APG+restart algorithm is given in Appendix \ref{ap:apg}. Additionally, we show that the traditional multiplicative update algorithm is extended for the weighted NMF in Appendix \ref{ap:multi}. The code for optimization is publicly available\footnote{{https://github.com/HajimeKawahara/sot}}. 

\section{Testing the Spin-Orbit Unmixing Using a Cloudless Toy Model}

\begin{figure}[]
 \begin{center}
   \includegraphics[width=\linewidth]{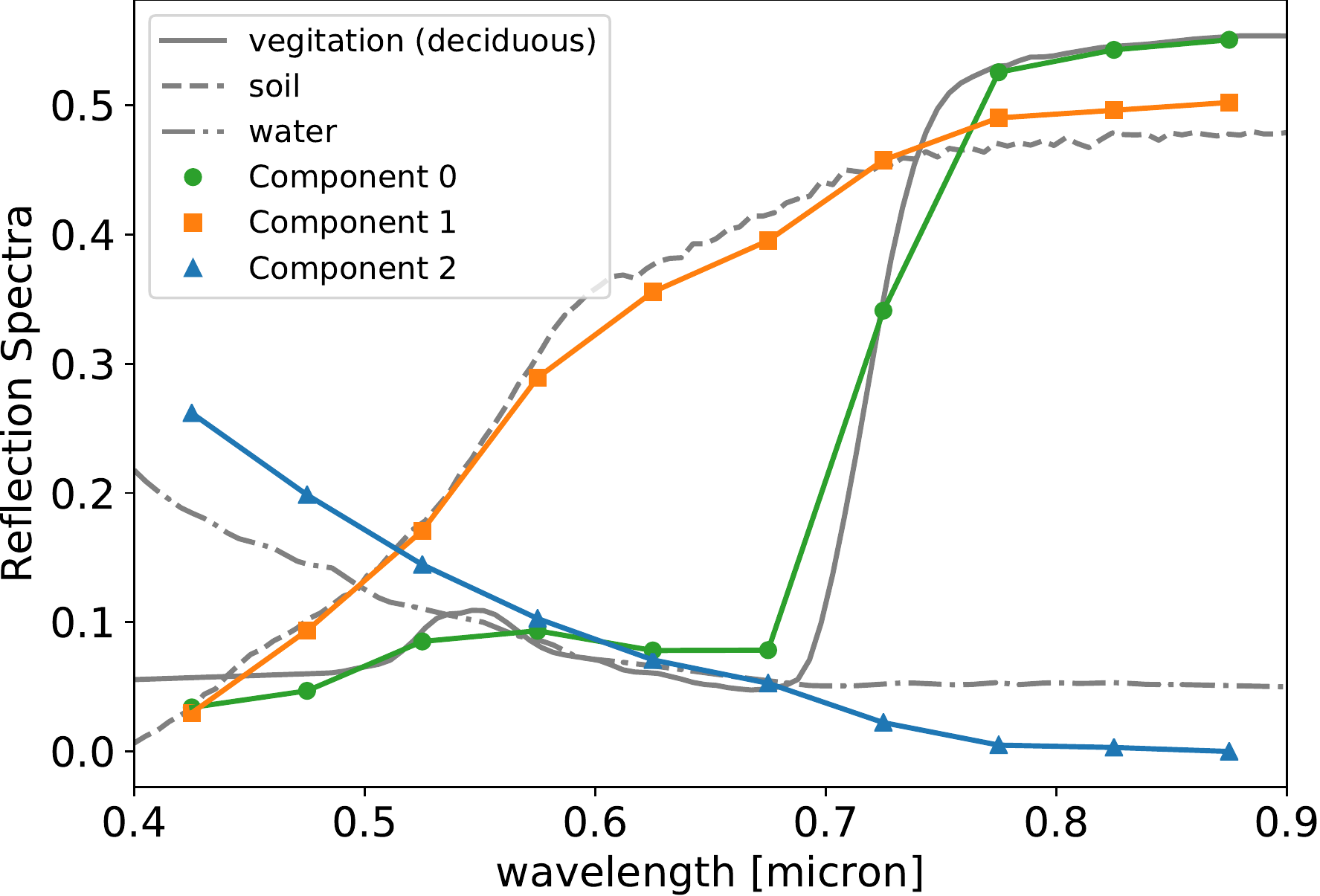}
 \end{center}
 \caption{Input (gray) and unmixed spectral components (color with markers) for the L2-VRDet model with $\lambda_A=10^{-1}$ and $\lambda_X = 10^{2}$. \label{fig:ref}}
\end{figure}

\begin{figure*}[]
 \begin{center}
   \includegraphics[width=0.49\linewidth]{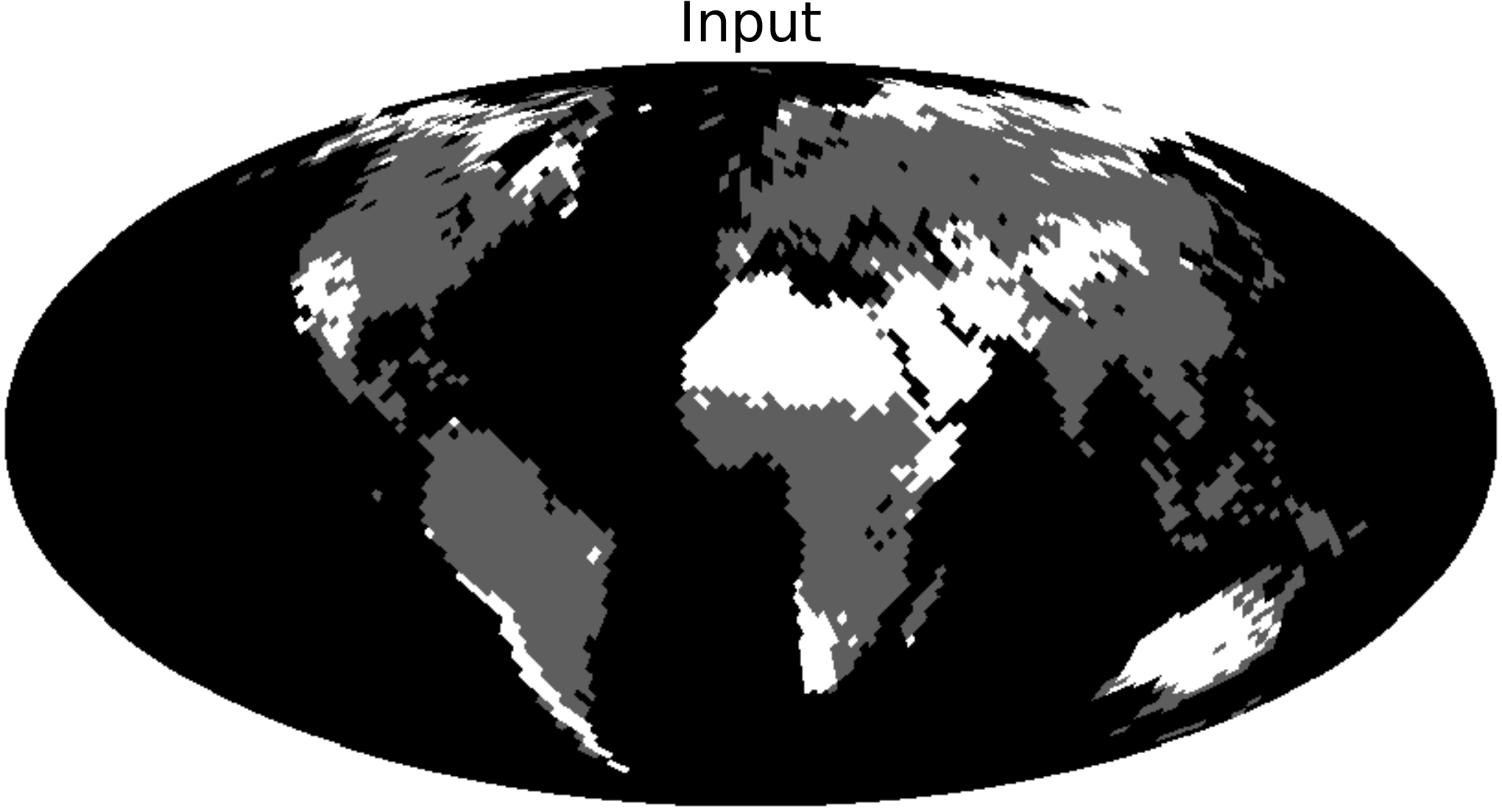}
         \includegraphics[width=0.49\linewidth]{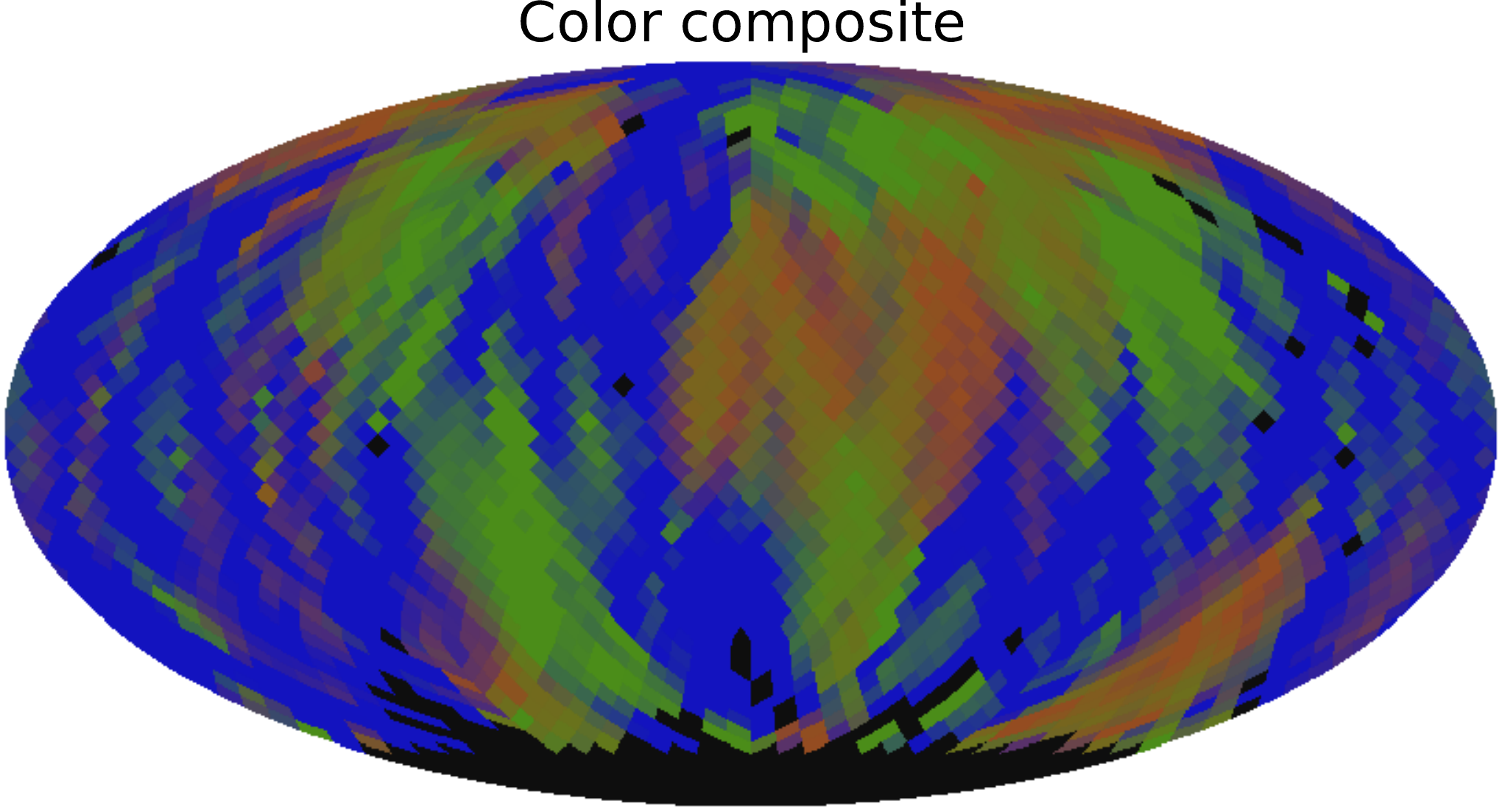}
 \end{center}
 \caption{Left: Input map of a toy model. The three colors indicate the different surface types, land, vegetation, and ocean, corresponding to white, gray, and black, respectively. Right: Color composite map for the same model. {\color{red} The color composite is based on the retrieved components in Figure \ref{fig:ref}; the components 0, 1, and 2 correspond to green, orange, and blue, respectively. } \label{fig:map}}
\end{figure*}

We test the spin-orbit unmixing by using the volume-regularized NMF and applying it to a toy model. The toy model assumes three surface types on a planet, including ocean, land, and vegetation. The reflection spectra for land and vegetation were taken from the ASTER spectral library \citep{baldridge2009aster}, and the ocean albedo is from \cite{1997JGR...10218801M}, as indicated by gray lines in Figure \ref{fig:ref}. The input classification of the map is based on the moderate resolution imaging spectroradiometer classification map in 2008, as shown in the left panel in figure \ref{fig:map}. We use the geometric settings of \cite{2012ApJ...755..101F}, an orbital inclination of 45$^\circ$, obliquity of $23.4^\circ$, and $\ThetaM=90^\circ$. We assume the spin rotation period is a sidereal day of Earth, 23.9344699/24.0 d and an orbital revolution period $P_\mathrm{orb}$ of 365 d. We took $N_i=$512 homogeneous samples over a year and injected a 1 \% Gaussian noise into the light curve. 

For the retrieval, we use a HEALPix map \citep{2005ApJ...622..759G} as $A_{jk}$ with $j=1,2..,N_\mathrm{pix}=3072$ pixels. In this test, we assume that we know the number of spectral components, $N_k=3$. Furthermore, we assume that we know the axial tilt parameters ${\bf g}$ and set $10^5$ as the number of iterations for the optimization. 

\begin{figure*}[]
 \begin{center}
   \includegraphics[width=0.32\linewidth]{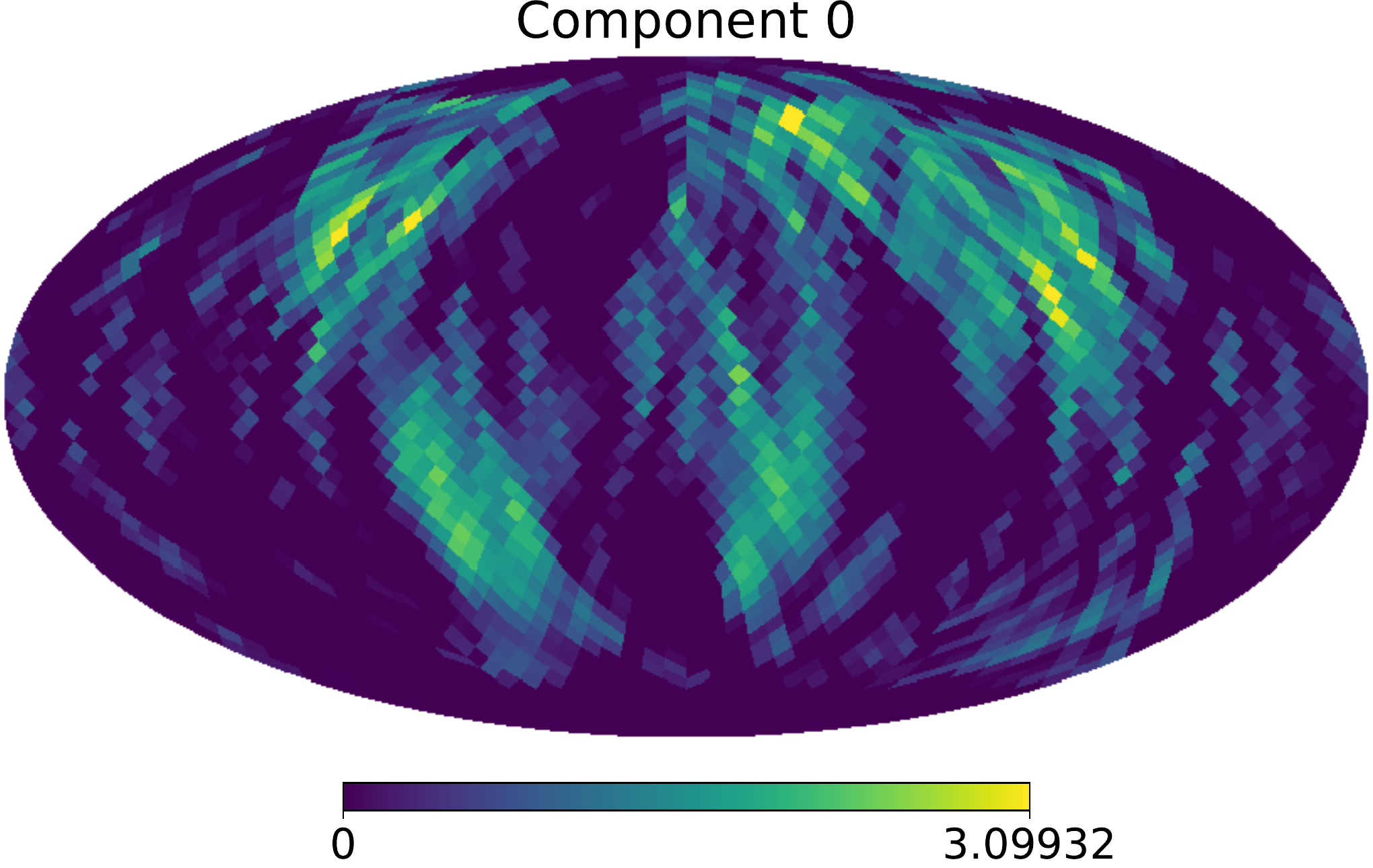}
   \includegraphics[width=0.32\linewidth]{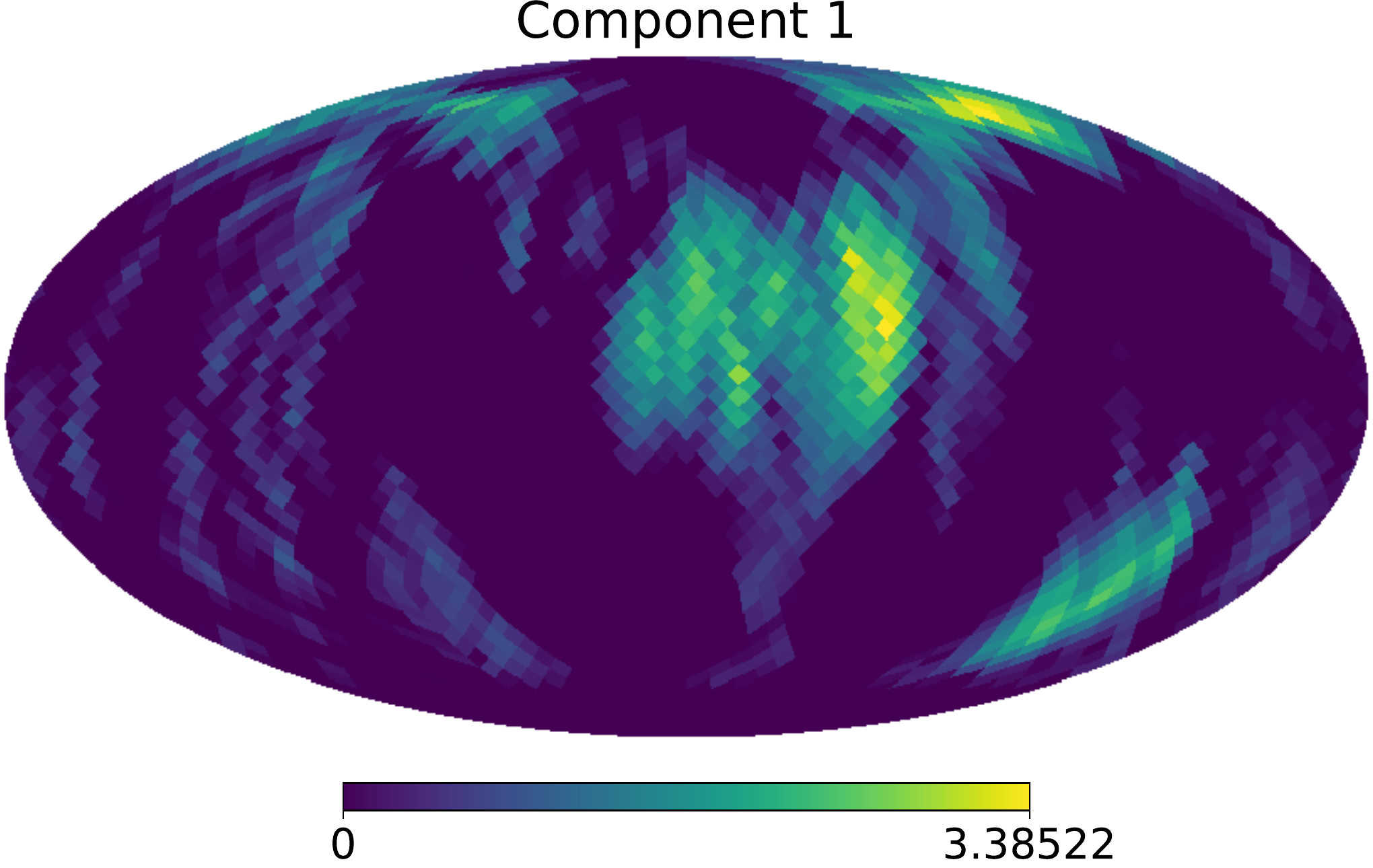}
   \includegraphics[width=0.32\linewidth]{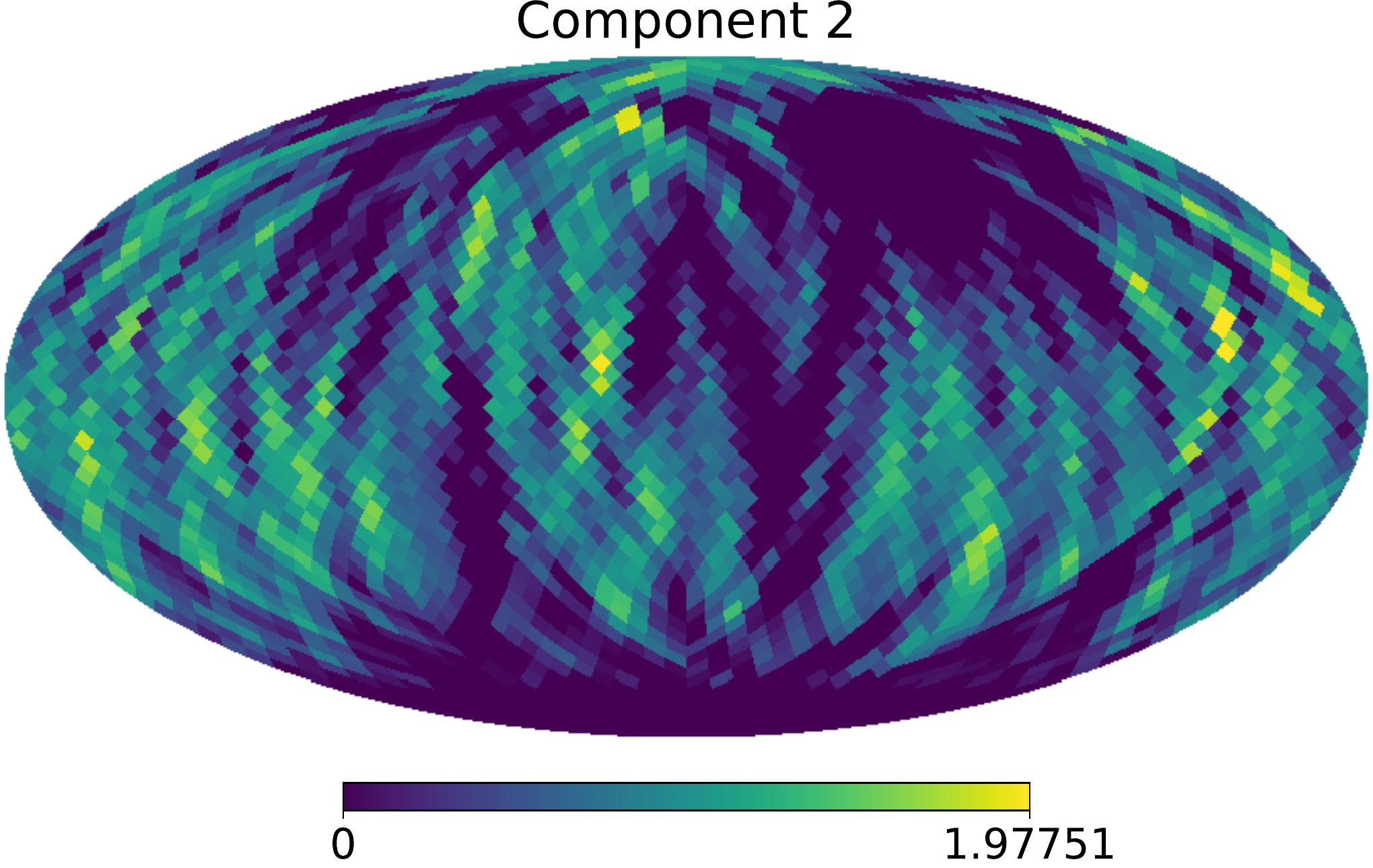}
 \end{center}
 \caption{Retrieved maps for different unmixed components 0, 1, and 2 from {\color{red} left to right}. We adopt L2-VRDet model with $\lambda_A=10^{-1}$ and $\lambda_X = 10^{2}$. \label{fig:C}}
\end{figure*}

Figures \ref{fig:ref} and \ref{fig:C} are examples of unmixed spectra and retrieved maps for the L2-VRDet model ($\lambda_A=10^{-1}$ and $\lambda_X = 10^{2}$). Because the normalization of each component is arbitrary, we adjust the normalization of each component to the input spectra. In this case, the input spectra and geography are accurately reproduced by the unmixed spectra and their retrieved distributions of components 0, 1, and 2, which corresponds to vegetation, land, and water, respectively. These results indicate that the spin-orbit unmixing using the volume-regularized NMF can infer the spectral components and their geography simultaneously. 

The sparsity of the retrieved maps is a notable feature of the spin-orbit unmixing that utilizes NMF. Because of the non-negative constraint, large parts of the maps remain zero. This feature was observed in the spin-orbit tomography using BVLS in \cite{2010ApJ...720.1333K}. In contrast, the spin-orbit tomography that uses the Tikhonov regularization does not exhibit such sparsity \citep{2011ApJ...739L..62K}.

We made a color composite map from the three maps in Figure \ref{fig:C}, as shown in the right panel of Figure \ref{fig:map}. These results show that the L2-VRDet model with an appropriate regularization can infer a global composition map for the toy model.

\begin{figure}[]
 \begin{center}
   \includegraphics[width=\linewidth]{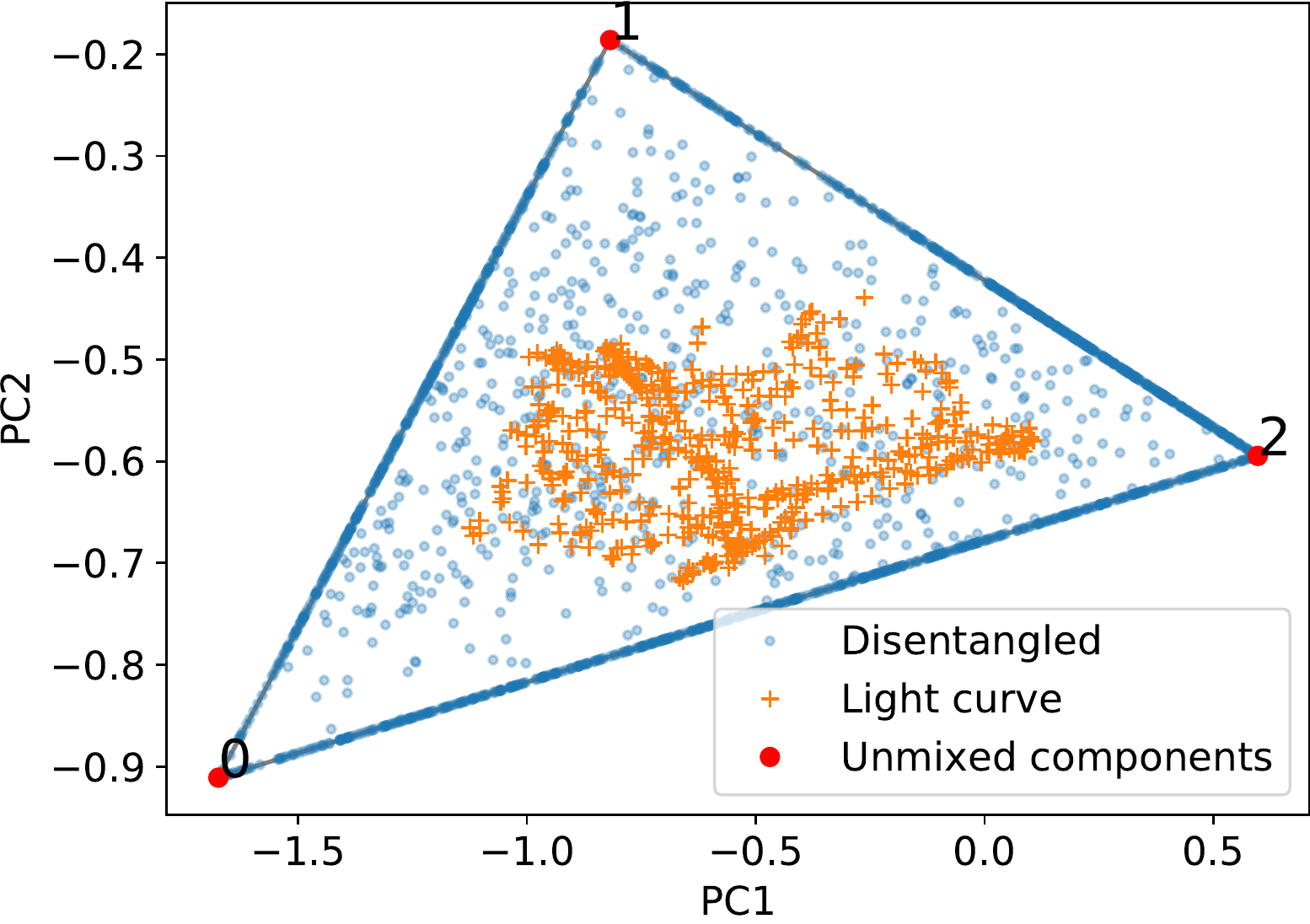}
 \end{center}
 \caption{Input light curve (orange cross), unmixed spectral components (red points), disentangled spectra (blue dots) on the PC1 -- PC2 plane. A simplex defined by the components 0, 1, and 2 are shown by the gray triangle. \label{fig:pca}}
\end{figure}

{\color{red} We remind the reader that the spectra are well mixed even for the cloudless toy model.} To illustrate how the spectra are unmixed, we project the input light curves and unmixed spectra as end members onto the PC1 -- PC2 plane in Figure \ref{fig:pca}. The PCA is computed using the input light curve. To draw this plot, we first compute the normalized light curve via the mixing matrix $\tilde{A} \equiv W A$, that is 
\begin{eqnarray}
\label{eq:direct}
D = \tilde{A} \, X. 
\end{eqnarray}
The light curve is normalized as $\tilde{D}_{il} = D_{il}/\sum_k \tilde{A}_{ik} $. We then derive PC1 and PC2 using $\tilde{D}_{il}$. The projection of $\tilde{D}_{il}$ and $X$ onto the PC1--PC2 plane, indicated by the orange crosses and red points, was computed using equation (\ref{eq:PCAproj}). The light curve does not touch the boundary of the triangle, which is defined by end members, that is, the triangle is not a convex hull of the light curve. This is because the spectra of the light curve are well mixed and the purity is low. In this case, the geometric disentanglement is essential for the spectra unmixing because the endmembers are far from the trajectory of the light curves.  The ``disentangled spectra'' of the light curve are defined by $\tilde{X} \equiv A X $, that is  
\begin{eqnarray}
D = W \tilde{X}.
\end{eqnarray}
The blue dots are the projection of the disentangled spectra normalized by $\sum_k A_{jk}$ onto the PC1--PC2 plane. The disentangled spectra are well spread in the triangle, and therefore, the triangle defines a convex hull of the disentangled spectra. The effect of the geometric disentanglement of the spectral unmixing is visualized as the expansion from orange crosses to blue dots in Figure \ref{fig:pca}.    

\subsection{Dependence on Regularization Parameters}

\begin{figure}[]
 \begin{center}
   \includegraphics[width=\linewidth]{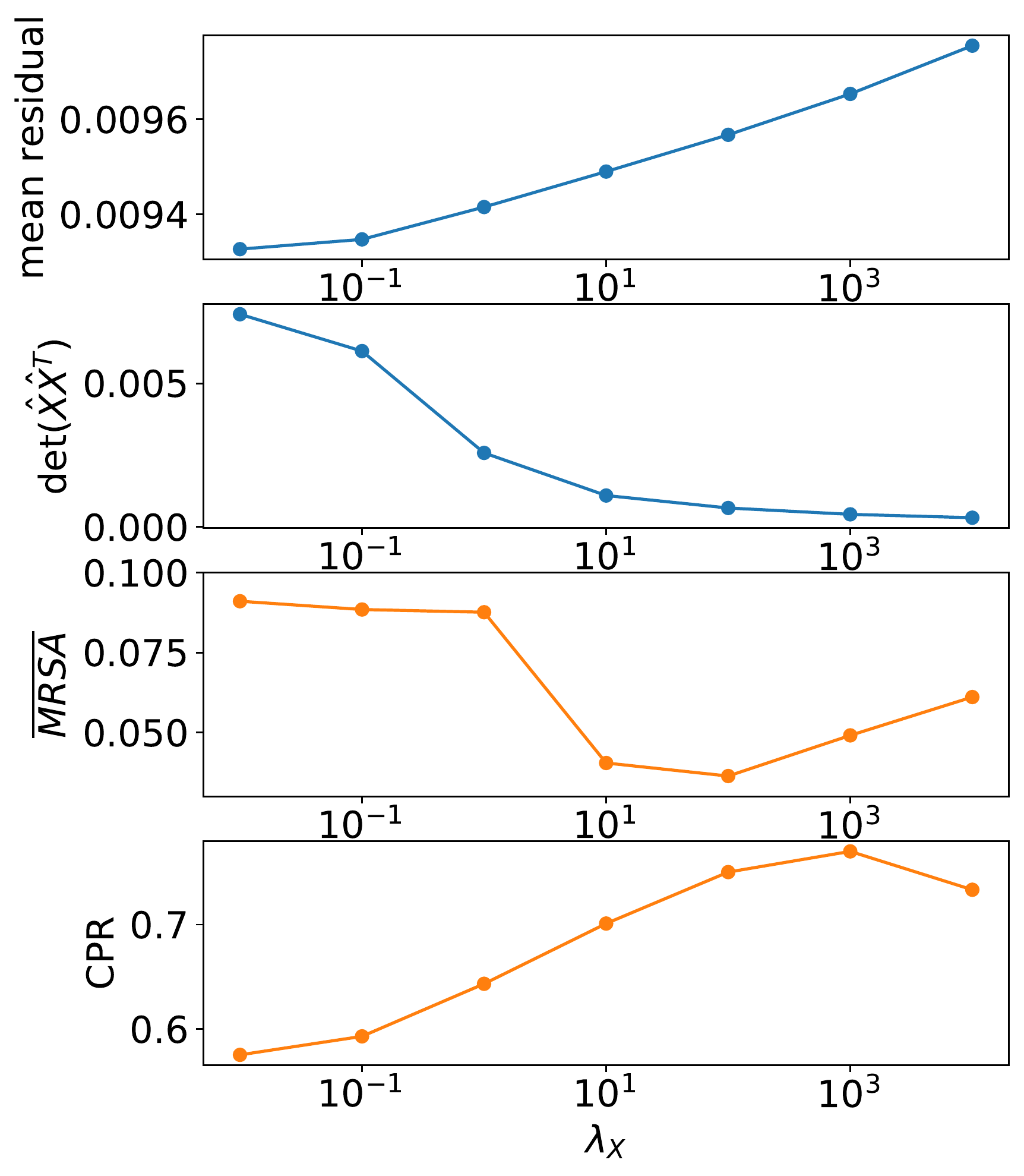}
 \end{center}
 \caption{The residuals, the surrogate of the normalized spectral volume, mean MSRA, and CPR as a function of $\lambda_X$ from top to bottom. {\color{red} We fix $\lambda_A = 10^{-1}$ in these panels.}  
 \label{fig:regx}}
\end{figure}

The over-regularization of the simplex volume (i.e. large $\lambda_X$) induces a worse fit of the data. This is confirmed by the mean residual of fitting the model to the data.  
\begin{eqnarray}
\mathrm{mean\,residual} \equiv \frac{1}{\overline{D}}\sqrt{\frac{||D - W A X||_F^2}{N_l N_i}},
\end{eqnarray}
where $\overline{D}$ is the mean value of the data. The top panel of Figure \ref{fig:regx} presents the mean residual as a function of $\lambda_X$. The mean residual gradually increases as the spectral regularization parameter increases. {\color{red}$\lambda_X = 10^{-2}$ and $10^{-1}$ }have similar mean residuals, which indicates that the model fits the data well for small regularization parameters. A smaller spectral regularization provides fewer constraints on the simplex volume minimization. Hence, there is a trade-off relation between the model’s goodness of fit and volume minimization. 

The panel below the top one illustrates a surrogate of the spectral of volume of normalized spectral components $\det{(\hat{X} \hat{X}^T)}$,
where $\hat{X}_{kl} = X_{kl}/\sum_l {X_{kl}}$. This quantity decreases as the spectral regularization parameter increases, which indicates that the spectral volume is minimized more as $\lambda_X$ increases. The surrogate of the normalized spectral volume roughly converges at $\lambda_X = 10^{1}$

\begin{figure}[]
 \begin{center}
   \includegraphics[width=0.95\linewidth]{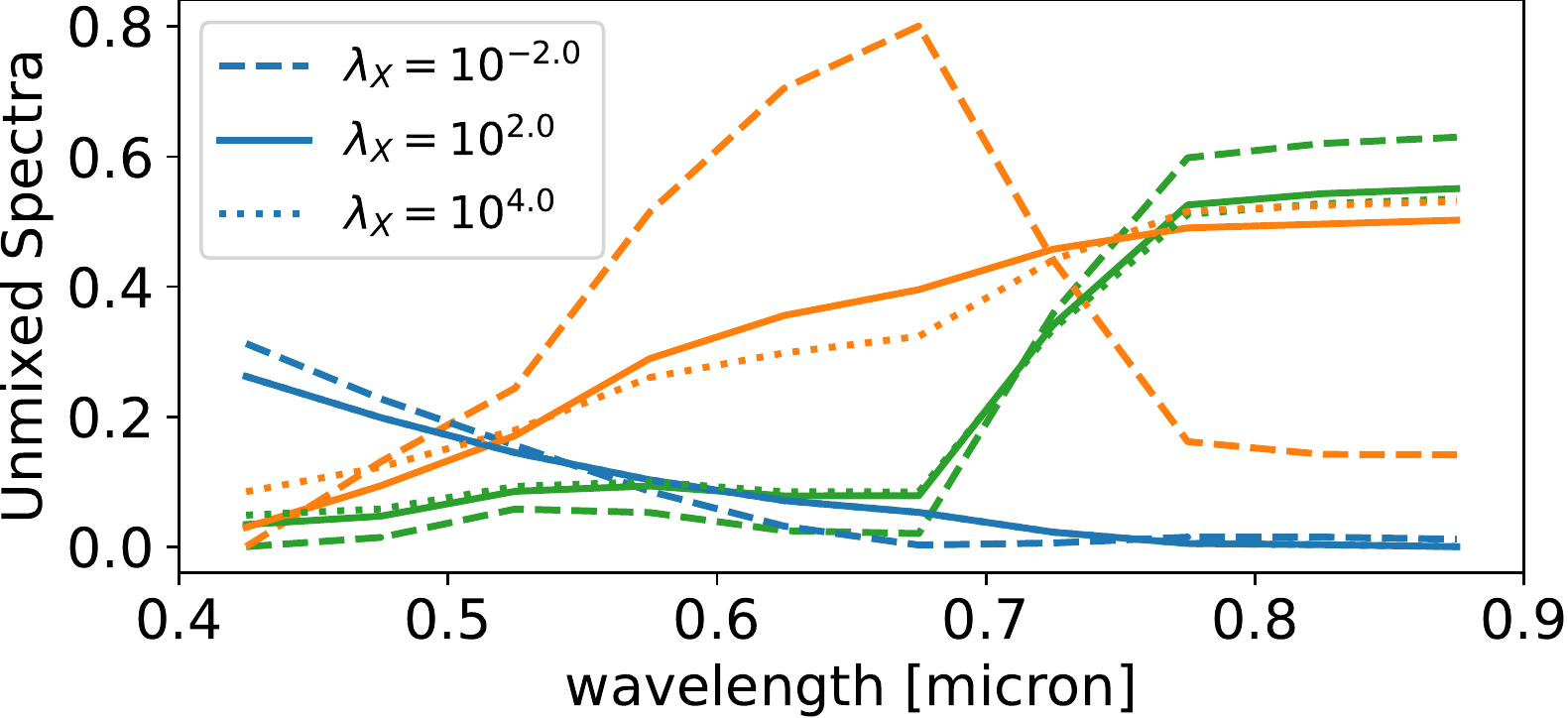}
 \end{center}
 \caption{Unmixed spectra for $\lambda_X=10^{-2},10^2$, and $10^{4}$. The color is the same as Figure \ref{fig:ref}. {\color{red} We fix $\lambda_A = 10^{-1}$ in these panels. }   \label{fig:regxf}}
\end{figure}

A direct comparison with the ground truth is useful to see how $\lambda_X$ affects the estimate of the spectral components and geography. To quantify the difference between the unmixed spectra and ground truth, we define the mean removed spectral angle (MRSA) between the two vectors $\xv$ and $\yv$ as 
\begin{eqnarray}
\displaystyle{\mathrm{MRSA}(\xv,\yv) =  \frac{1}{\pi} \cos^{-1} \left( \frac{(\xv - \overline{\xv})^T (\yv - \overline{\yv})}{||\xv - \overline{\xv}||_2 ||\yv - \overline{\yv}||_2} \right),}
\end{eqnarray}
where $\overline{\xv}$ and $\overline{\yv}$ are the mean of $\xv$ and $\yv$ respectively, and MRSA $(\xv, \yv) \in [0,1]$. The two vectors perfectly match when the MRSA $(\xv, \yv) =0$. The third panel shows the mean of MRSA over the components 
\begin{eqnarray}
\overline{\mathrm{MRSA}} =  \sum_k \mathrm{MRSA}(\xv_k,{\xv}_k^{\mathcal{G}})/N_k,
\end{eqnarray}
where ${\xv}_k^{\mathcal{G}}$ is the ground truth (input spectrum). Moreover, figure \ref{fig:regxf} shows the actual shapes of the unmixed spectra for different $\lambda_X$. The difference is observed for Component 2 (orange, land). The unmixed spectra when $\lambda_X \le 10^0$ resulted in a worse fit to the ground truth. Interestingly, when $\lambda_X \ge 10^3$, the fit of the unmixed spectra to the ground truth also worsened. This is possibly due to the over-constraint on the spectral model which in turn restricts its ability to explain the data accurately (large residuals); this results in an incomplete estimate of the spectral components. 

The comparison between the retrieved map and ground truth is quantified by the Correct Pixel Rate (CPR), which is defined by the correct answer rate of the classification map. The bottom panel of figure \ref{fig:regx} shows its dependence on $\lambda_X$. Insufficient spectral regularization resulted in not only a worse mean MRSA, but also a worse estimate of the geography. In regards to both the mean MRSA and CPR, an optimal range for $\lambda_X$ of $10^{1} $--$ 10^{3}$ was observed. 

Although we cannot compute the MRSA and CPR for unknown geographies and surface spectra, these results suggest that a curve of a surrogate of normalized spectral volume, as a function of $\lambda_X$, can be used to determine the optimal value of $\lambda_X$. We suggest the following procedure: (1) Plot the mean residual and $\det{(\hat{X} \hat{X}^T)}$ as a function of $\lambda_X$. (2) Observe the change of the spectral shape as a function of $\lambda_X$. (3) Use $\lambda_X$ at a turning point of the spectral shape and $\det{(\hat{X} \hat{X}^T)}$ and avoid a large value for the mean residual. 

\begin{figure}[]
 \begin{center}
   \includegraphics[width=\linewidth]{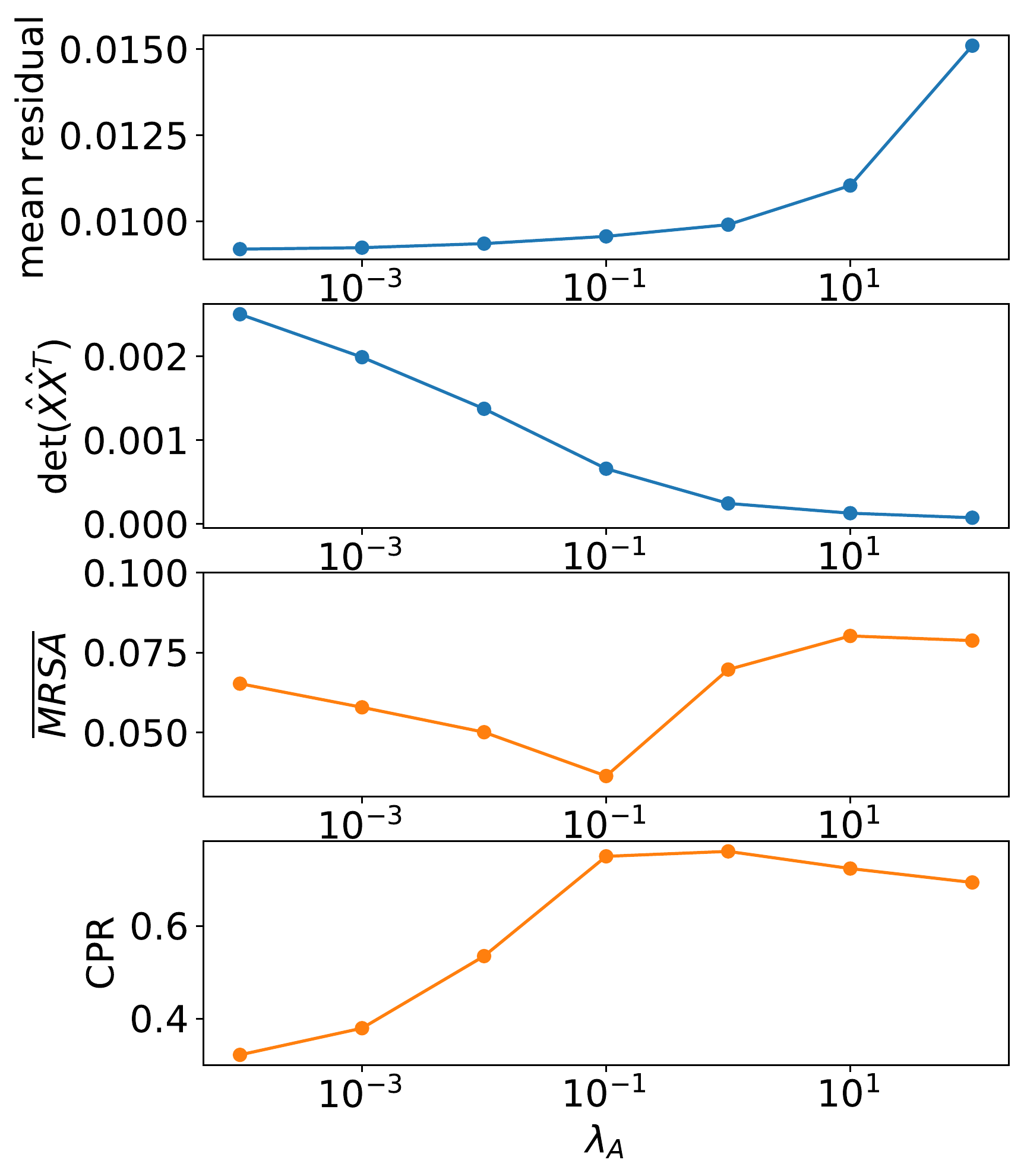}
 \end{center}
 \caption{Mean residual, a surrogate of the normalized spectral volume, mean MSRA and CPR as a function of $\lambda_A$ from top to bottom. {\color{red} We fix $\lambda_X=10^2$ in these panels. }.    \label{fig:rega}}
\end{figure}

\begin{figure}[]
 \begin{center}
\includegraphics[width=0.8\linewidth]{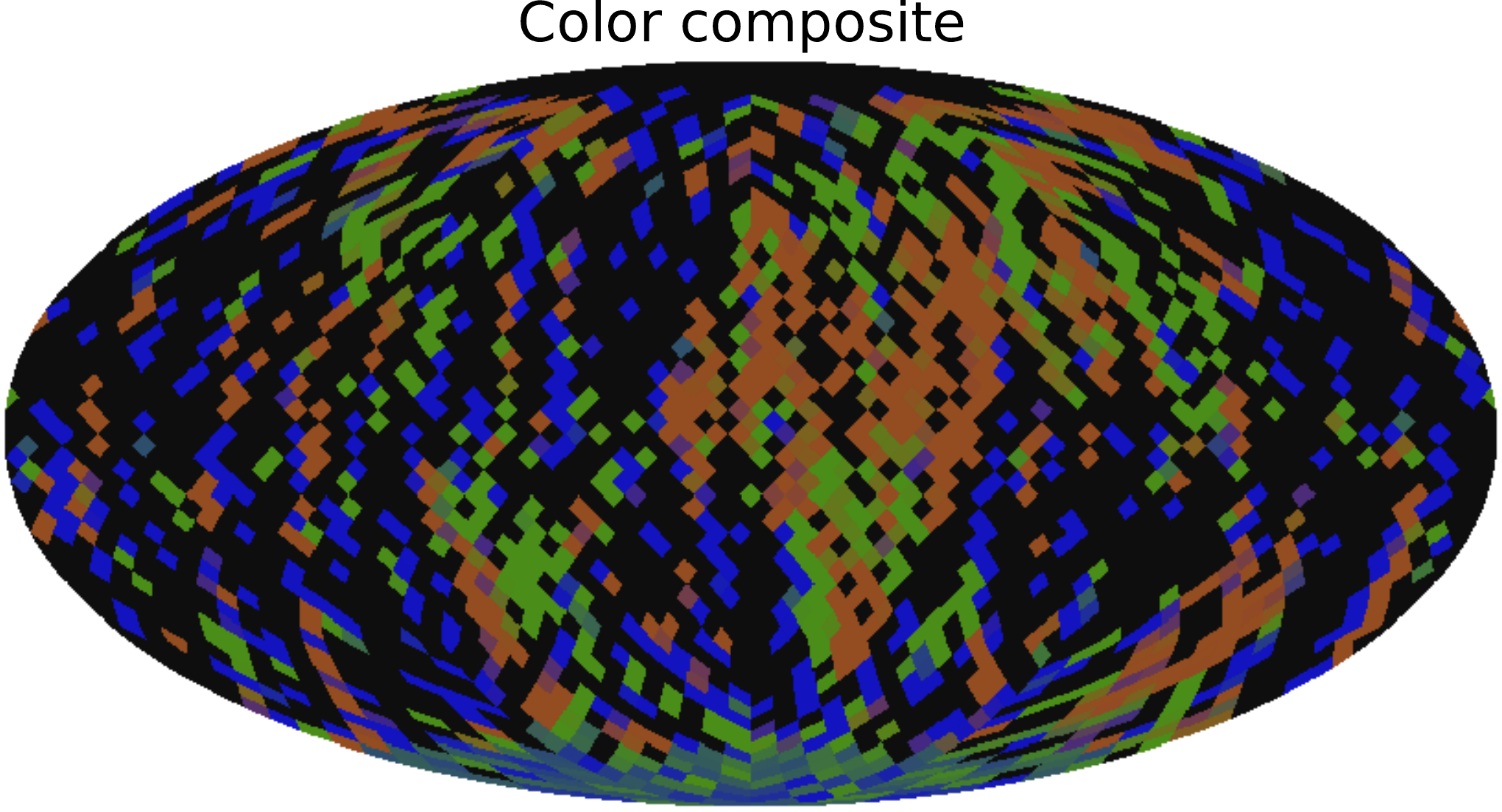}
 \end{center}
 \caption{Example of the color composite map for insufficient spatial regularization ( $\lambda_A=10^{-3}$). We adopt $\lambda_X = 10^{2}$ {\color{red} to make this figure}. \label{fig:regaf}}
\end{figure}

Figure \ref{fig:rega} shows similar plots to those in Figure \ref{fig:regx}, but for the spatial regularization $\lambda_A$. The over-constraint on $A$ (i.e. large $\lambda_A$) contributes to a bad fit of the data and a smaller volume of spectral components. Smaller spatial regularization parameters resulted in a bad estimate of the geography, as indicated by the CPR. This is because insufficient spatial regularization creates a noisy map due to the instability of the mapping (see Figure \ref{fig:regaf} as an example).  Contrastingly, a large $\lambda_A$ will slightly decreases the CPR because it will imply that the inferred spectra are getting worse, and the spatial resolution of the map is decreasing in quality. This poor resolution will result in a large mean residual. {\color{red}Therefore, one should check both the residual and the surrogate of normalized spectral volume as a function of $\lambda_A$ because these quantities have a trade-off relation. We suggest choosing the optimal $\lambda_A$ as the smallest value that (1) keeps the noise in the inferred map nonsignificant and (2) avoids a large mean residual.}

We have described how spatial and spectral regularization parameters affect the results; we have also discussed a guideline to follow when choosing the optimal spatial and spectral parameters. To find the optimal $\lambda_X$ (or $\lambda_A$), {\color{red} we fixed $\lambda_A$ (or $\lambda_X$) in Figure \ref{fig:regx} (or Figure \ref{fig:rega}) }. In practice, this procedure should be iterative so that we can find the optimal set of $\lambda_A$ and $\lambda_X$. We recognize that our current guideline for choosing the optimal parameters is not quantitative. Ideally, the performance of the prediction can be used to choose the optimal parameters, such as cross-validation. However, the large computational time of the optimization method is too long to perform a cross-validation. Therefore, we postpone the quantitative criterion needed to choose the parameters for further study.

\subsection{Choice of the Number of Spectral Components}
So far, we have assumed the number of spectral components $N_k = 3$. Generally, $N_k$ should be one of the free parameters. Here, we consider the cases for when $N_k=2$ (over-constrained) and $N_k=4$ (under-constrained). The cost function for $N_k=2$ ($Q=6 \times 10^4$) is much larger than that of $N_k=3$ ($Q=$2355) and 4 ($Q=$2565). This indicates that $N_k=2$ is insufficient to explain the data. For $N_k=4$, we could not reach the convergence of the cost function {\color{red} (\ref{eq:cost}) with the regularization term (\ref{eq:vrnmfreg})} even though  the number of iterations reached $10^6$ which was where we stopped. 

Although information criteria such as the Akaike Information Criterion (AIC) are used as the model selection for a different number of free parameters, the degrees of freedom are not clear for the inverse problem. Ignoring this fact, if we evaluate AIC by $-2 \log{\mathrm{(Likelihood)}} + 2 \mathrm{(degree\,of\,freedom)} = ||D-WAX||_F^2/\sigma^2 + 2 N_k N_j$, where $\sigma$ is the standard deviation of the input noise, we obtain AIC = 135158.8, 23141.4, and 28889.3 for $N_k=2,3,$ and 4, respectively. These results might indicate that $N_k=3$ is the optimal number for the components. 

Another problem in real data is that it is often difficult to estimate the likelihood because we do not understand the statistical nature of the noise. In this case, the cross validation is often used as the model selection. However, the cross validation is unrealistic because of the high computational cost of the current scheme. We postpone the criterion that will allow us to choose the optimal number of surface components for further study. Hence, in this paper, we require the number of surface components as prior knowledge for mapping.

{\color{red}
\subsection{Comparison with Spectral Unmixing on Light Curves}
So far, we have explained how geography is disentangled from spectra in spin-orbit unmixing. Here, we consider spectral unmixing on the light curve with no disentanglement of geometry and compare it with the unified model. By minimizing the cost function
\begin{eqnarray}
\label{eq:WAXdirect}
Q = \frac{1}{2}|| D -  \tilde{A} X ||_F^2 + \frac{\lambda_X}{2} \det{(X X^T)}  \\
\hbox{ \,\, subject to } \tilde{A}_{ik} \ge 0, X_{kl} \ge 0,
\end{eqnarray}
we obtain the unmixed spectral components for different spectral regularization as shown in Figure \ref{fig:refdirect}. We find that both components 0 and 1, which can be interpreted as surface components on continents, are sensitive to the volume regularization. These results show that the NMF with simplex volume minimization works even without geometric disentanglement.

Compared with the spin-orbit unmixing, the spectrum of soil (gray dashed) is less reproduced by the component 1 (orange) even for the best case, $\lambda_X = 10^{0}$ and the results are more sensitive to the choice of $\lambda_X$ (see Figure \ref{fig:regxf} for comparison).  This is likely because the geometric disentanglement is essential to sufficiently separate the spectrum of soil from that of vegetation. 

\begin{figure}[]
 \begin{center}
\includegraphics[width=\linewidth]{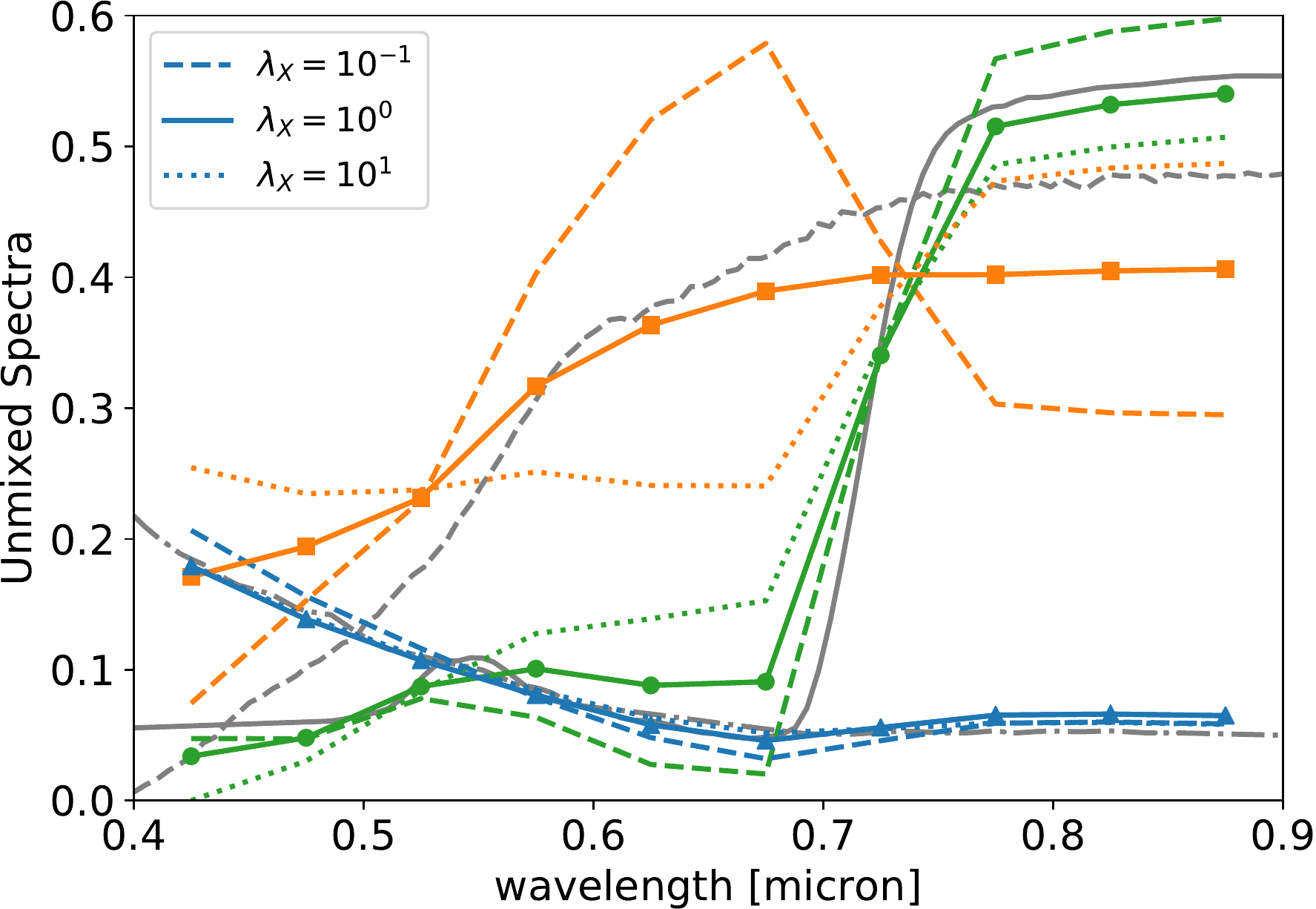}
 \end{center}
 \caption{  Unmixed spectral components (colors with markers) for the direct spectral unmixing of the light curve with $\lambda_X=10^{-1}, 10^{0},$ and $ 10^{1}$. The green circles, orange squares, and blue triangles correspond to components 0, 1, and 2, respectively. The gray lines are input spectra the same as those in Figure \ref{fig:ref}. \label{fig:refdirect}}
\end{figure}
}

\section{Application to DSCOVR Data}

In this section, we demonstrate our method using real multiband light curves of the Earth as observed by DSCOVR \citep{2018AJ....156...26J}.  DSCOVR has been continuously monitoring our Earth from the L1 point since 2015. The geometry provided by DSCOVR is not the same as the geometry provided by direct imaging, because DSCOVR continuously looks almost at the {\color{red} dayside} of Earth. However, the geometric kernel contains latitudinal information because of the axial tilt of the Earth. This enables us to do a two-dimensional mapping \citep{2019ApJ...882L...1F}. We use seven optical bands (0.388, 0.443, 0.552, 0.680, 0.688, 0.764, and 0.779 $\mu$m) in DSCOVR filters ($N_l=7$). The band widths are very narrow (0.8 -- 3.9 nm) and there are strong oxygen B and A absorption in 0.688, 0.764 $\mu$m. Owing to computational efficiency, we use one-fourth of the two-year data (i.e. one in each four bins) used in \cite{2019ApJ...882L...1F}, resulting in a number of $N_i=2435$ time bins.

\begin{figure}[]
 \begin{center}
\includegraphics[width=\linewidth]{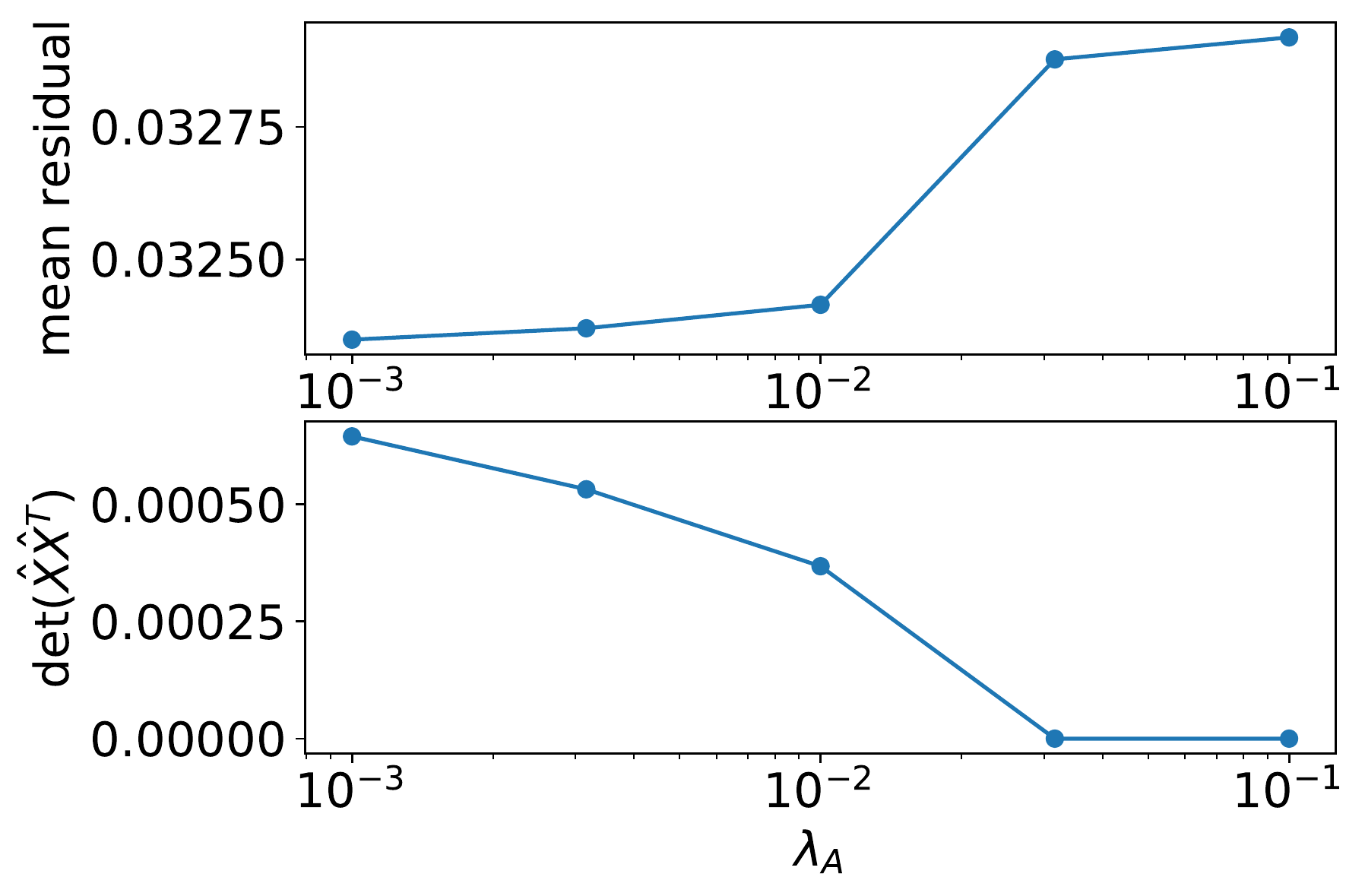}
\includegraphics[width=\linewidth]{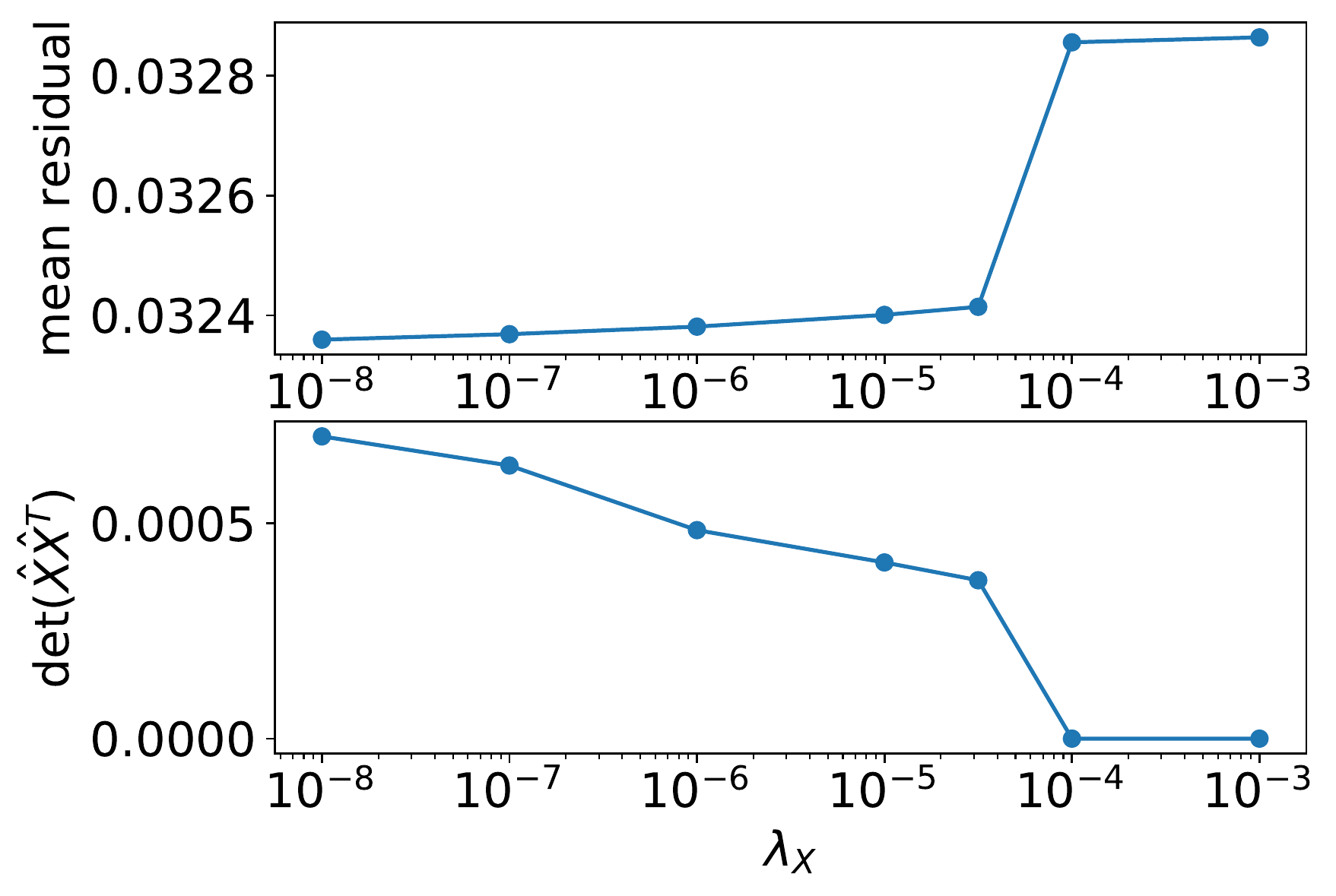}
 \end{center}
 \caption{Mean residual and the surrogate of the normalized spectral volume as functions of $\lambda_A$ (top; $\lambda_X = 10^{-4.5}$) and  $\lambda_X$ (bottom; $\lambda_A = 10^{-2}$). We take $10^{-2}$ as the optimal value of $\lambda_A$ because of a significant increase at $\lambda_A=10^{-1.5}$. Also, we take $10^{-4.5}$ as the optimal value of $\lambda_X$  because of a significant increase {\color{red} in the mean residual} at $\lambda_X=10^{-4}$. \label{fig:regdd}}
\end{figure}

\begin{figure}[]
 \begin{center}
\includegraphics[width=\linewidth]{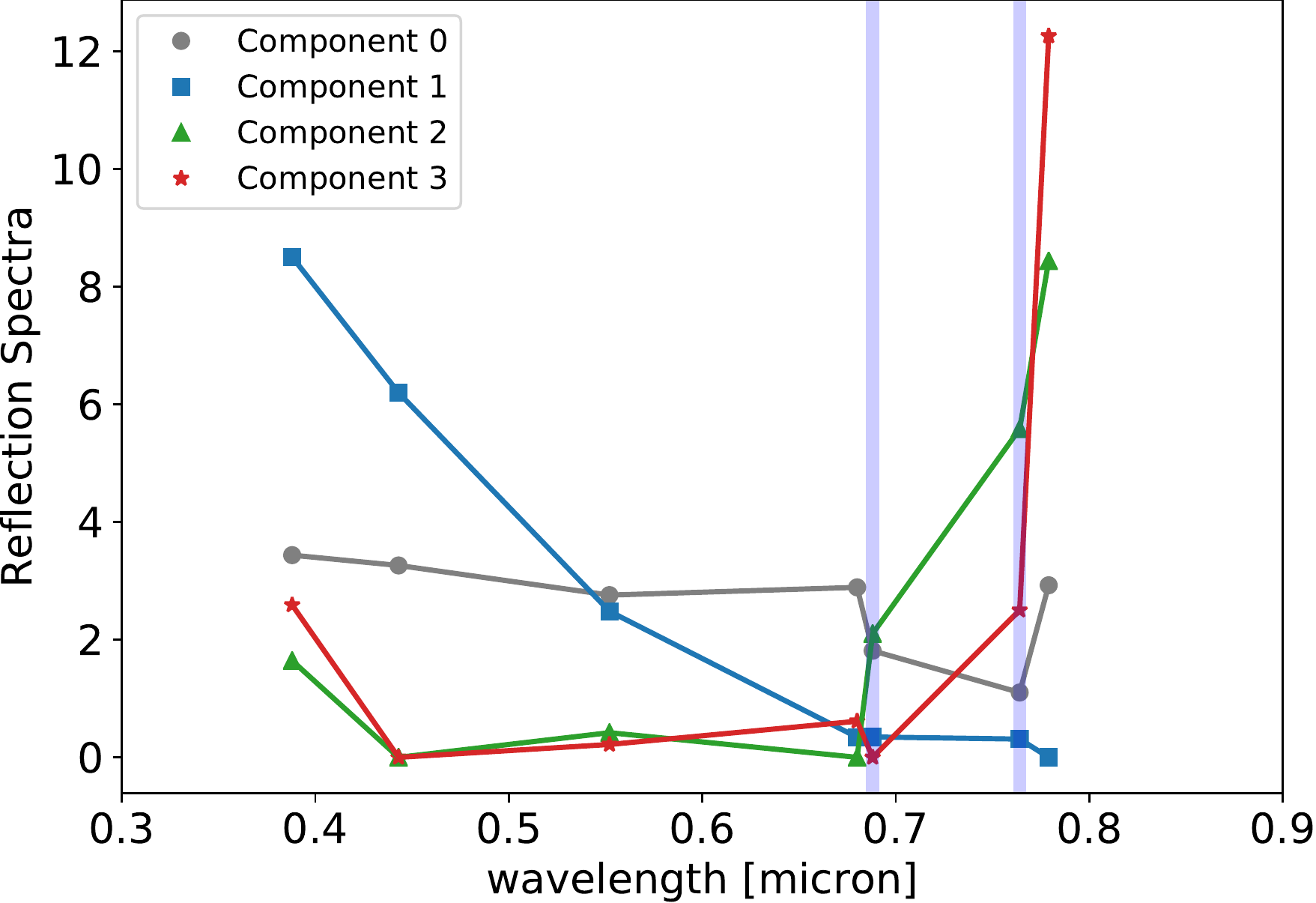}
\includegraphics[width=\linewidth]{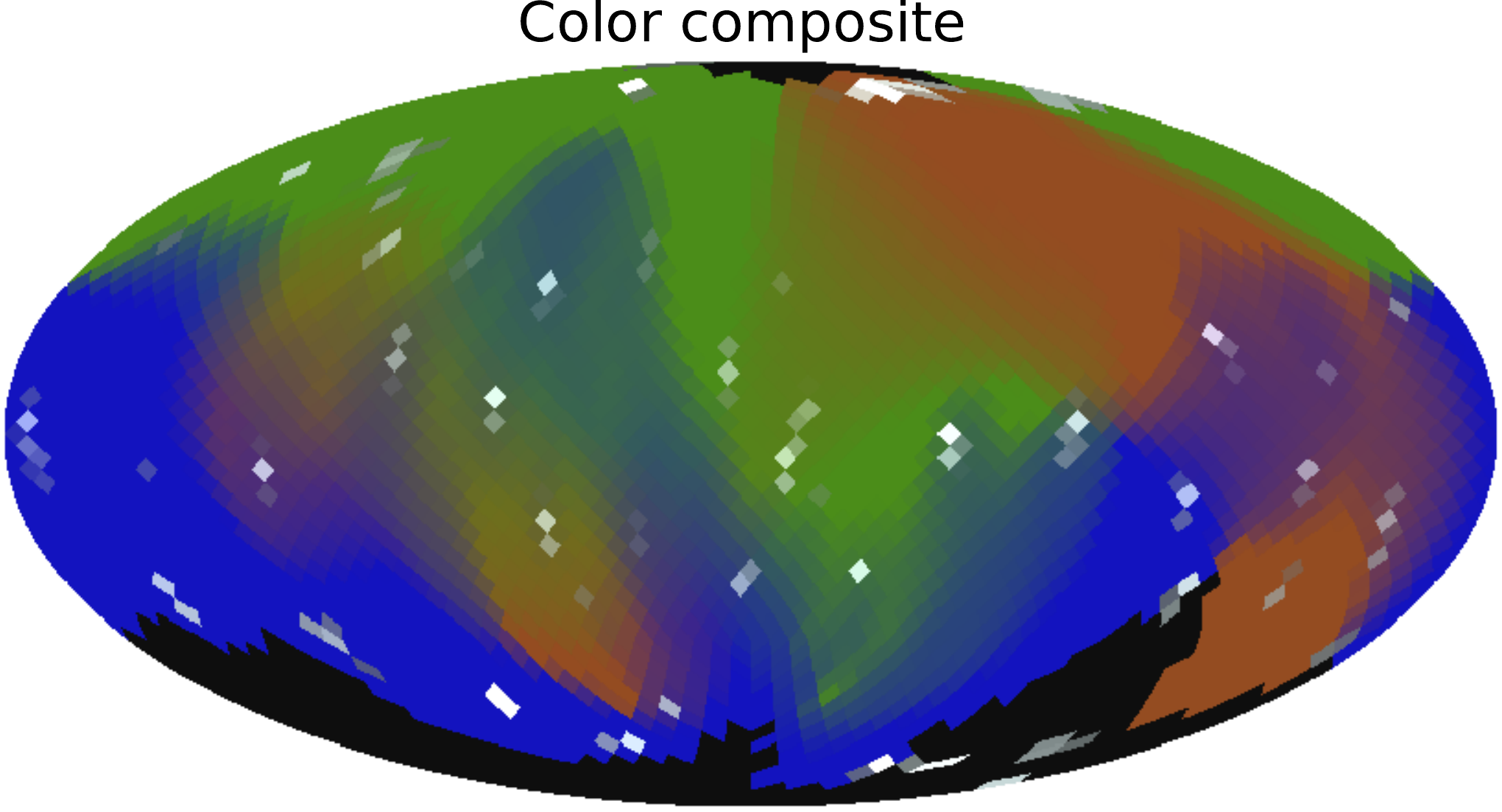}
 \end{center}
 \caption{{\color{red} Normalized} unmixed spectra (top) and color composite map (bottom) for the DSCOVR data.  In the top panel, both 0.688 and 0.764 $\mu$m bands are strongly affected by oxygen absorption (shaded by blue vertical lines).  The bottom panel shows a color composite map. We use white, blue, green, and brown for components 0, 1, 2, and 3, respectively.  \label{fig:dscovr}}
\end{figure}

Figure \ref{fig:dscovr} shows the unmixed spectra and color composite map when we assume that $N_k = 4$. For regularization parameters, we followed the procedure described in the previous section. Figure \ref{fig:regdd} shows the mean MRSA and surrogate of the normalized spectral volume. It was observed that $\lambda_X=10^{-4.5}$ and $\lambda_A=10^{-2}$ are the optimal values, because a significant increase of the mean residual is observed at the range larger than these values. Component 1 accurately reproduced the actual geography and blue spectrum of the ocean; components 2 and 3 reproduced the continent distribution of Earth. {\color {red} Component 1 is less sensitive to the choice of $\lambda_X$ compared with components 2 and 3, therefore, component 1 is a relatively robust estimate of a surface component.}  From the unmixed spectrum, component 2 resembled the spectrum of vegetation because of the increase larger than 0.688 micron although, the strong oxygen absorption at 0.688 and 0.764 microns suppressed this increase to some extent. Component 3 corresponds to the spectrum of soil or sands. In fact, the continent of Australia  (less vegetation) was painted by component 3. The southern part of Africa and the Amazon (large forest areas) are roughly painted by component 2. We do not have enough spatial resolution around North Africa, Eurasia, and Europe. Although we did not consider our scheme being able to clearly distinguish between soil and vegetation, we believe that the differences between components 2 and 3 reflect the variety of spectra of land continents on planet Earth.  

Component 0 exhibited a flat spectrum except for strong oxygen absorption bands (0.688, 0.764 $\mu$m) reproducing the cloud or ice spectrum. {\color {red} Component 0 as well as component 1 are less sensitive to the choice of $\lambda_X$ compared with components 2 and 3.} However, the distribution of component 0 is patchy, except for the localization at the North Pole.  The patchy distribution probably reflects a temporal cloud distribution because real clouds do not have a static distribution, some of which might be from the ice near the pole. 

{\color{red} These patchy pixels have values that are roughly $\alpha=5$ times larger compared to those of other components. Also, the unmixed spectrum of component 0 is $\beta=50$ times higher on wavelength average than the total value of those of other components. The fraction of the patchy pixels is about $\gamma \sim 1/100$. Multiplying $\alpha$, $\beta$, and $\gamma$, we find that the power of component 0 in the patchy pixels is roughly several times higher than the total power of other components. This value is consistent with the contribution of clouds on reflected light on Earth. The fact that the cloud component is localized in these patchy pixels represents a limitation of the current method, which assumes that all of the components have a static distribution over the observation period.} Further improvement is needed so that non-static components can be included to the model.

\section{Summary and Discussion}

In this paper, we constructed a unified retrieval model for  spectral unmixing and spin-orbit tomography (spin-orbit unmixing) using the nonnegative matrix factorization and L2 and volume regularization. The spin-orbit unmixing works on the cloudless toy model and real multicolor light curves by DSCOVR. Here, we raise several remaining issues that we did not consider in this study.

The simultaneous estimate of the axial tilt parameters ${\bf g}$ is first. For simple two-dimensional mapping, \cite{2016MNRAS.457..926S} analyzed how the axial tilt parameters are inferred from amplitude modulation, and \cite{2018AJ....156..146F} constructed a Bayesian framework to estimate the parameters. Similar work should also be done in the spin-orbit unmixing. However, the computational cost will be an issue that would need to be addressed, as the optimization of {\color{red}NMF} requires a high numerical cost. 

Moreover, the clock setting problem still remains. Thus far, all of the works done on two-dimensional mapping assume that we know the exact phase of the geometric kernel. The spin rotation period, derived by the auto-correlation function, was assumed to be used \citep{2012ApJ...755..101F}. However, as \cite{2016ApJ...822..112K} pointed out, the apparent periodicity of the photometric variability is not identical to the spin rotation period. The frequency modulation analysis provides the spin rotation period. For instance, we need to check if the spin-orbit unmixing works well when we use an inferred spin rotation period from the frequency modulation. Otherwise, a technique with a simultaneous estimate of the spin might be required. 

The next challenge is how  non-static compositions such as clouds can be included in the model {\color{red} \citep[see][as an attempt of the time-dependent mapping]{2019arXiv190312182L}}. This will become vitally important when we apply this technique to gaseous planets. 

Another challenge is the dependency on results of various types of regularization. For instance, \cite{Aizawa} reported that the L1+TSV regularization provided better results than the Tikhonov regularization. Furthermore, several other types of volume regularization have been proposed in the field of remote sensing \citep[e.g.][]{2019arXiv190304362A}; therefore, a comparative study of regularization is required. 

Additionally, a more quantitative criterion is needed to select the optimal number of surface components  and the regularization parameters. Because of the high computational cost, the cross validation is unrealistic for the current scheme. An objective criterion to select these parameters will help us to apply the technique to unknown exoplanets where the ground truth is not known. 

The author is grateful to the DSCOVR team for making the data publicly available. I deeply appreciate Siteng Fan and Yuk L. Yung for providing the processed light curves and their geometric kernel from the DSCOVR dataset. I would also like to thank Masataka Aizawa, Kento Masuda, {\color{red} Nick Cowan} for their insightful discussions. {\color{red} I would also like to thank the anonymous reviewer for a careful reading and constructive suggestions.} This work was supported by JSPS KAKENHI grant Numbers JP17K14246, JP18H04577, JP18H01247, and JP20H00170. This work was also supported by the JSPS Core-to-Core Program Planet2 and SATELLITE Research from Astrobiology center (AB022006).

%
%

\appendix
\section{Geometric Kernel of the Spin-Orbit Tomography}\label{ap:sot}

\subsection{Disk-integrated Scattered Light}

\begin{figure*}[]
 \begin{center}
   \includegraphics[width=0.35\linewidth]{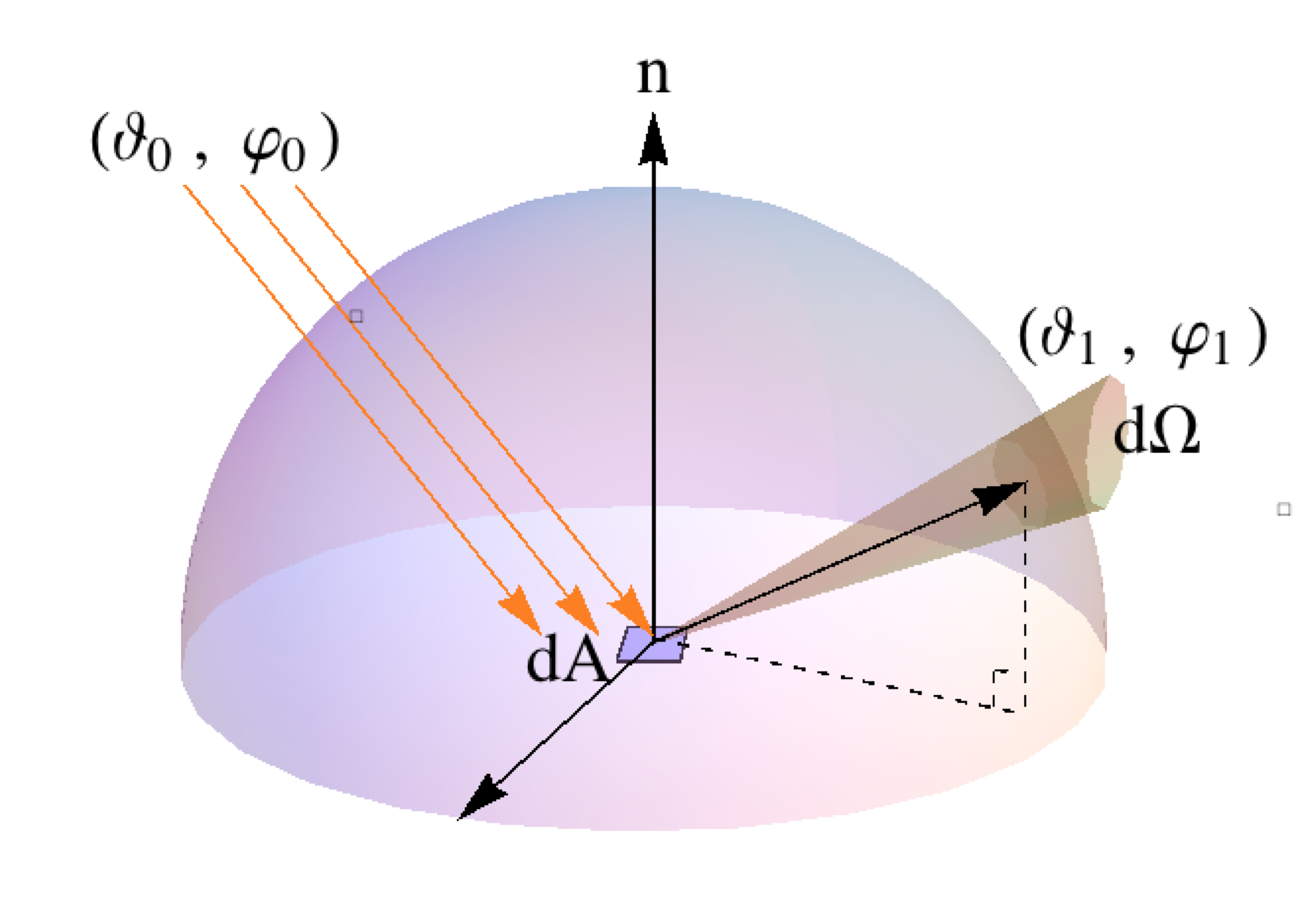}
   \includegraphics[width=0.35\linewidth]{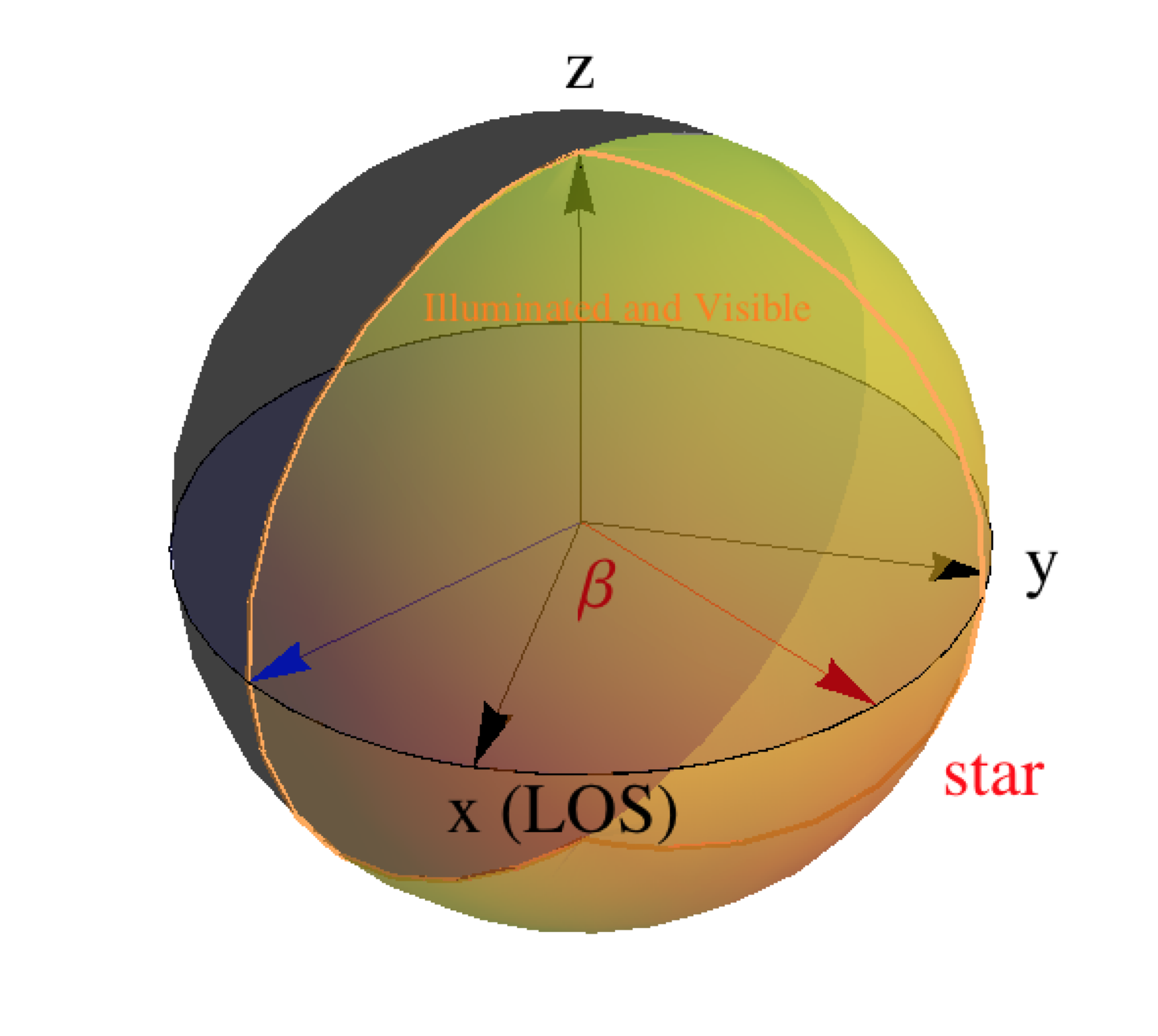}
 \end{center}
 \caption{Left: Incoming light and outcoming light of a small facet $d A$ on the surface of a planet. Right: The visible and illuminated (IV) region of a planet surrounded by the orange curve. \label{fig:geo}}
\end{figure*}

Here, we summarize the computation of the reflection light from a planet to an observer. The outward energy from a facet $d A$ to a direction with a solid angle $d \Omega$ (the left panel in Figure \ref{fig:geo}) is expressed as
\begin{eqnarray}
d E = L_{\uparrow} \cos{\vartheta_1} d A d \Omega d\lambda,
\end{eqnarray}
where $L_{\uparrow}$ is the upward radiance, and ${\vartheta_1}$ is a zenith angle between a direction and a normal vector. Let us assume that we observe flux from a planet at a distance of $d$ using a telescope with an effective area $A_\mathrm{tel}$, then light in a cone with a solid angle $d \Omega = d A_\mathrm{tel}/d^2$ contributes to the flux. Therefore, the flux from a facet $d A$ on a plane to the telescope area $d A_\mathrm{tel}$ can be written as 
\begin{eqnarray}
\label{eq:brdfdef3}
\Delta E d A_\mathrm{tel} &=& L_\uparrow \cos{\vartheta_1} d \Omega d A = \frac{L_\uparrow}{d^2} \cos{\vartheta_1} d A d A_\mathrm{tel}.
\end{eqnarray}
Thus, we obtain the total flux from a planet as 
\begin{eqnarray}
\label{eq:brdfdef6x}
f_p &=& \int_\mathrm{planet} \Delta E = \int_\mathrm{planet} d A \frac{L_\uparrow}{d^2} \cos{\vartheta_1}.
\end{eqnarray}

The BRDF of the surface element $s$ is defined by the ratio of the outward radiance to the inward irradiance,
\begin{eqnarray}
\label{eq:brdfdef}
R^s(\vartheta_0,\varphi_0,\vartheta_1,\varphi_1) \equiv \pi \frac{\mathrm{L_\uparrow (\vartheta_1, \varphi_1)}}{E_\downarrow (\vartheta_0, \varphi_0)}.
\end{eqnarray}
where $\vartheta_0$ and $\varphi_0$ are the solar zenith angle and azimuth angle, respectively, and $\varphi_1$ is the azimuth angle to an observer (see Figure \ref{fig:geo})\footnote{We inserted a factor of $\pi$ so that the BRDF becomes identical to the reflectivity when the scattering is isotropic.}. For most surface types, the BRDF almost solely depends on a relative azimuth angle $\varphi = \varphi_1 - \varphi_0$ instead of each azimuth angle as
\begin{eqnarray}
\label{eq:brdfdef22}
  R^s(\vartheta_0,\varphi_0,\vartheta_1,\varphi_1) = R^s(\vartheta_0,\vartheta_1,\varphi).
\end{eqnarray}
The stellar irradiance is expressed as
\begin{eqnarray}
\label{eq:brdfdef4}
E_\downarrow (\vartheta_0) = \frac{L_\star}{4 \pi a^2} \cos{\vartheta_0} = \frac{f_\star d^2}{a^2} \cos{\vartheta_0},
\end{eqnarray}
where $a$ is the star-planet distance, and ${L_\star}$ and ${f_\star}$ are the stellar luminosity and flux, respectively. The flux from a planet is expressed as
\begin{eqnarray}
\label{eq:brdfdef4x}
f_p &=& \int_\mathrm{IV} d A \frac{E_\downarrow (\vartheta_0)}{\pi d^2}  R^s(\vartheta_0,\vartheta_1,\varphi) \cos{\vartheta_1} \nonumber \\
&=&  \frac{f_\star R_p^2}{\pi a^2} \int_{\mathrm{IV}} d \Omega_1  R^s(\vartheta_0,\vartheta_1,\varphi)  \cos{\vartheta_0} \cos{\vartheta_1},
\end{eqnarray}
where IV is the illuminated and visible region as shown in the right panel of figure \ref{fig:geo}.

Assuming an isotropic reflection $R^s(\vartheta_0,\vartheta_1,\varphi)  = m (\theta, \phi) $, we obtain,
\begin{eqnarray}
f_p &=& \int d \Omega_1  W_{\bf g}(t,\theta,\phi) m(\theta, \phi)  
\end{eqnarray}
where $W_{\bf g}(t,\theta,\phi)$ is the geometric kernel for the Lambert approximation.
\begin{eqnarray}
W_{\bf g}(t,\theta,\phi) = 
  \left\{
    \begin{array}{l}
\displaystyle{\frac{f_\star R_p^2}{\pi a^2}  \cos{\vartheta_0} \cos{\vartheta_1} \mbox{ for $\cos{\vartheta_0}, \cos{\vartheta_1}>0$}}\\
\\
\displaystyle{ 0 \mbox{\,\, otherwise,} }
    \end{array}
  \right.
\end{eqnarray}

Here, we define the three fundamental vectors, $\eS, \eO,$ and $\eR $ which are the unit vector from the planet center to the stellar center, from the planet center to the observer, and the normal unit vector at the planet surface, respectively. Using them, we can rewrite $\cos{\vartheta_0} = \eS \cdot \eR$ and $\cos{\vartheta_1} = \eO \cdot \eR$. Using the orbital phase $\Theta$ and an orbital inclination $i$, we obtain
\begin{eqnarray}
  \label{eq:eS}
\eS &=& (\cos{(\Theta-\ThetaM)}, \sin{(\Theta-\ThetaM)}, 0)^T,\\
  \label{eq:eO}
\eO &=& (\sin{i} \cos{\ThetaM}, -\sin{i} \sin{\ThetaM}, \cos{i})^T,
\end{eqnarray}
where $\ThetaM$ is the orbital phase at equinox. 

 We also define the spherical coordinate fixed on the planet surface,
\begin{eqnarray}
\label{eq:vectsphe}
\eRd  (\phi, \theta) = (\cphi \stheta, \sphi \stheta, \ctheta)^T.
\end{eqnarray}

Applying a spin rotation along $\Phi$ and a rotation matrix $\mathcal{R} (\zeta)$ as a function of a planet’s obliquity $\zeta$, we get
\begin{eqnarray}
  \label{eq:eR}
\eR &=& \mathcal{R} (\zeta)  \, \eRd (\phi+\Phi, \theta) \nonumber \\
&=& \left(
\begin{array}{c}
\cos{(\phi+\Phi)} \stheta \\
\czeta \sin{(\phi+\Phi)} \stheta +  \szeta  \ctheta \\
- \szeta \sin{(\phi+\Phi)} \stheta + \czeta \ctheta
\end{array} \right).
\end{eqnarray}
The geometric weight is given by 
\begin{eqnarray}
\label{eq:geoweight}
W_{\bf g}(t,\theta,\phi) &=& 
  \left\{
    \begin{array}{l}
\displaystyle{\frac{f_\star R_p^2}{\pi a^2} (\eS \cdot \eR) (\eR \cdot \eO) \mbox{ for $\eS \cdot \eR>0, \eR \cdot \eO>0$}}\\
\\
\displaystyle{ 0 \mbox{\,\, otherwise.} }
    \end{array}
  \right. 
\end{eqnarray}

In addition, we consider the case where the reflectivity is constant and isotropic over the surface, $R^s(\vartheta_0,\vartheta_1,\varphi)  = R$, (the Lambert approximation). We take $\eO = (1,0,0)^T$ and define the phase angle $\beta = \eS \cdot \eO$, i.e. $\eS = (\cos{\beta}, \sin{\beta}, 0)^T$. These definitions yield equation (\ref{eq:brdfdef4x}): 
\begin{eqnarray}
\label{eq:brdfdef4xCC}
f_p &=&  \frac{f_\star R_p^2 R}{\pi a^2} \int_{-\pi/2 + \beta}^{\pi/2} d \phi \int_0^{\pi} d\theta \sin^2{\theta} \cos{\phi} ( \cos{\beta} \cos{\phi} \sin{\theta} + \sin{\beta} \sin{\phi} \sin{\theta} ) \\
&=& \frac{2 R}{3} \phi_p(\beta) \left( \frac{R_p}{a} \right)^2 f_\star,
\end{eqnarray}
where 
\begin{eqnarray}
\label{eq:brdfdef4xCCCX}
\phi_p(\beta) \equiv \frac{1}{\pi} [\sin{\beta} + (\pi - \beta) \cos{\beta}],
\end{eqnarray}
is the Lambert phase function.

\section{Optimization of the Weighted NMF by a Block Coordinate Descent}

The block coordinate descent \citep[e.g.][]{kim2014algorithms,zhou2011minimum, 2019arXiv190304362A} consists of the following two subproblems:
\begin{itemize}
    \item QP(A): optimization of a quadratic form for $\av_k$ (the {\color{red} column} vector of $A$) 
    \item QP(X): optimization {\color{red} of} a quadratic form for $\xv_k$ (the {\color{red} row} vector of $X$)
\end{itemize}
The block coordinate descent solves these quadratic problems (QP(A) and QP(X)) iteratively using a nonnegative least square (NNLS) scheme. In this appendix, we derive the quadratic forms and then explain the projected gradient descent and its accelerated versions as the NNLS solver.

\subsection{Quadratic Programming}
\subsubsection*{Quadratic Form for $\av_k$}\label{ap:qp}
A (log) likelihood term for the weighted NMF can be rewritten in the quadratic form
\begin{eqnarray}
\label{eq:qfa}
\frac{1}{2} || D - W A X ||_F^2 &=&  \frac{1}{2} \sum_{i} \sum_{l} \left( \Delta_{il} - \sum_{j} W_{ij} A_{jk} X_{kl} \right)^2 \\
&=&  \frac{1}{2}  \sum_{i} \sum_{l} X_{kl} X^T_{lk}\left( \sum_{j} W_{ij} A_{jk} \right)^2 - \sum_{i} \sum_{l} \Delta^T_{li} \sum_{j} W_{ij} A_{jk} X_{kl} + \frac{1}{2} ||\Delta||_F^2\\
\label{eq:FQP}
&=& \frac{1}{2} \sum_l X_{kl}^2 \sum_{j^\prime, j} A^T_{k j^\prime} \left( \sum_{i} W^T_{j^\prime i} W_{ij} \right) A_{jk} - \sum_j \left[ \sum_{l} X_{kl} \left(\sum_{i} \Delta^T_{li} W_{ij} \right) \right] A_{jk}  + \frac{1}{2} ||\Delta||_F^2\\
&=& \frac{1}{2} \av^T_k \mathcal{L}_A \av_k - \lv_A^T \av_k + \mathrm{const.}
\end{eqnarray}
where 
\begin{eqnarray}
\mathcal{L}_A &\equiv& \xv_k^T \xv_k  W^T W  \\
\lv_A &\equiv&  W^T \Delta \, \xv_k,
\end{eqnarray}
and  $\Delta = \Delta (k)$ is defined by $\Delta_{il} \equiv  D_{il} - \sum_{s \neq k} \sum_{j} W_{ij} A_{js} X_{sl}$. The penalty of the Tikhonov regularization (L2 term) is 
\begin{eqnarray}
\frac{1}{2} \lambda_A || A ||_F^2 &=& \frac{1}{2} \av^T_k \mathcal{T}_A \av_k + \mathrm{const.} \\
\mathcal{T}_A &\equiv& \lambda_A I
\end{eqnarray}
Thus, the quadratic programming for the weighted NMF with a spatial Tikhonov regularization minimizes  
\begin{eqnarray}
q_A &=& \frac{1}{2} \av^T_k (\mathcal{L}_A + \mathcal{T}_A) \av_k - \lv_A^T \av_k .
\end{eqnarray}

\subsubsection*{Quadratic Form for $\xv_k$}

Likewise, we obtain the (log) likelihood term as a quadratic form of $\xv_k$ from equation (\ref{eq:FQP}) as,
\begin{eqnarray}
\label{eq:qfx}
 \frac{1}{2}  || D - W A X ||_F^2 &=&  \frac{1}{2} \xv^T_k \mathcal{L}_X \xv_k - \lv_X^T \xv_k + \mathrm{const.} \\
 \mathcal{L}_X &\equiv& ||W \av_k||_2^2 \, I \\ 
 \lv_X &\equiv& \Delta^T W \av_k 
\end{eqnarray}

The volume regularization of the Gram determinant term can be written in the quadratic form of $\xv_k$
\begin{eqnarray}
\label{eq:detxxt}
 \frac{1}{2} \lambda_X \det{(X X^T)} &=& \frac{1}{2} \xv_k^T \mathcal{D}_X  \xv_k \\
\mathcal{D}_X  &\equiv& \lambda_X \det{(\breve{X}_k \breve{X}_k^T)} \left[ I - \breve{X}_k^T  (\breve{X}_k \breve{X}_k^T)^{-1} \breve{X}_k  \right], 
\end{eqnarray} 
where $\breve{X}_k$ is a submatrix of $X$ when we remove the $k$-th row of $X$. The derivation of equation (\ref{eq:detxxt}) is given in \cite{zhou2011minimum}.

In Table \ref{tab:quadratic}, we summarize the quadratic terms for different regularization types. This list also includes the log-determinant type of the volume regularization \cite{ang2018volume} and a simple L2 term for $\xv_k$. 

\begin{table*}[]
	\centering
	\caption{Terms in quadratic problems. }
	\begin{tabular}{ccll}
		\hline\hline
		Term & Cost function ($\times 2$) & $\mathcal{W}_A$ or $\mathcal{W}_X$ & $\becv_A$ or $\becv_X$ \\
		\hline
		Likelihood for $\av_k$& $|| D - W A X ||_F^2$  &$\mathcal{L}_A = \xv_k^T \xv_k  W^T W$ & $\lv_A = W^T \Delta \, \xv_k$ \\
		Tikhonov (L2) term for $\av_k$& $\lambda_A ||A||_F^2$ & $\mathcal{T}_A = \lambda_A I_J$ & - \\ 
		Likelihood for $\xv_k$& $|| D - W A X ||_F^2$  &$\mathcal{L}_X = ||W \av_k||_2^2 \, I_L$ &
		$\lv_X = \Delta^T W \av_k $ \\ 
		Volume Regularization (Det) & $\lambda_X \det{(X X^T)}$ & $\mathcal{D}_X = \lambda_X \det{(\breve{X}_k \breve{X}_k^T)} [ I_L - \breve{X}_k^T  (\breve{X}_k \breve{X}_k^T)^{-1} \breve{X}_k ]$ & -  \\
		Volume Regularization (Logdet) & $\lambda_X \log{[\det{(X X^T+\delta I_K})]}$ & $\mathcal{E}_X = \lambda_X \mu_\mathrm{min}^{-1} I_L$ & -  \\
		Tikhonov (L2) term for $\xv_k$& $\lambda_X ||X||_F^2$ & $\mathcal{T}_X =\lambda_X I_L$ & - \\ 
	\end{tabular}
	\tablecomments{ $\breve{X}_k$ is a submatrix of $X$ by removing the $k$-th row of $X$, $\Delta_{il} =  D_{il} - \sum_{s \neq k} \sum_{j} W_{ij} A_{js} X_{sl}$, $I_J$ (or $I_L$, $I_K$) is an identity matrix $\in \mathbb{R}^{N_j \times N_j}$ (or $\mathbb{R}^{N_l \times N_l}$, $\mathbb{R}^{N_k \times N_k}$) , and $\delta$ is a small number (we adopt $10^
		{-6}$). The minimum eigenvalue of $E=\det{(X X^T+\delta I_K})$ is denoted by $\mu_\mathrm{min}$. In practice, we use $X$ in the previous iteration to compute $\mathcal{E}_X$ \citep{ang2018volume}. }
	\label{tab:quadratic}
\end{table*}

\subsection{Projected Gradient Descent}\label{ap:apg}

The projected gradient descent (PG)-based methods to solve a quadratic problem, 
\begin{eqnarray}
q = \xv^T \mathcal{W} \xv - \becv^T \xv.
\end{eqnarray} 
are described. The gradient descent with a nonnegative condition is given by 
\begin{eqnarray}
\xv^{(t+1)} = \mathcal{P}[\xv^{(t)} - \eta \nabla q] =  \mathcal{P}[\xv^{(t)} - \eta (\mathcal{W} \xv^{(t)} - \becv)],
\end{eqnarray} 
where the projection operator on a nonnegative orthant is defined by $ \mathcal{P}[\xv] = \{ \mathrm{max}(x_k, 0) \}$. We obtain the PG algorithm by adopting the inverse of the Lipschitz constant $L$ to $\eta$. As the Lipschitz constant, one can use the 2-norm of $ ||\mathcal{W}||_2 = \mathrm{max} (|| \mathcal{W} \xv ||_2/||\xv||_2)$ for $ \xv \in \mathbb{R}^m, \xv \neq 0$ or a Frobenius norm of $ ||\mathcal{W}||_F = \sqrt{\sum_j \sum_{i} \mathcal{W}_{ij}^2} = \sqrt{\mathrm{tr}(\mathcal{W}^T \mathcal{W})}$. Although the 2-norm is more efficient than the Frobenius norm (i.e. $||\mathcal{W}||_F \ge ||\mathcal{W}||_2$ ), the computational cost of the 2-norm is much higher than that of the Frobenius norm, especially for a large matrix\footnote{Therefore, we use a 2-norm for $X$ and a Frobenius norm for $A$.}.

\begin{algorithmic}
\STATE \underline{\bf Projected Gradient Descent (PG)}
\STATE Minimization of $ q = \xv^T \mathcal{W} \xv - \becv^T \xv$
\STATE Initialization: $T= I - \mathcal{W}/L, \sv = \becv/L, \xv_0$
\WHILE {Condition} 
        \STATE $\xv^{(t+1)} =  \mathcal{P}[T \xv^{(t)} + \sv]$
\ENDWHILE
\end{algorithmic}

The convergence rate of the PG algorithm is relatively slow. The PG algorithm with Nesterov's acceleration \citep{nesterov1983method} is called the accelerated projected gradient descent. The APG algorithm is summarized as follows. 

\begin{algorithmic}
\STATE \underline{\bf Accelerated Projected Gradient Descent (APG)}
\STATE Minimization of $ q = \xv^T \mathcal{W} \xv - \becv^T \xv$
\STATE Initialization: $T= I - \mathcal{W}/L, \sv = \becv/L, \xv^{(0)}, \yv^{(0)}=\xv^{(0)}, \alpha_0 = 0.9$ 
\WHILE {Condition} 
        \STATE $\xv^{(t+1)} =  \mathcal{P}[T \yv^{(t)} + \sv]$
        \STATE $\alpha_{t+1}=(\sqrt{\alpha_t^4 + 4\alpha_t^2} - \alpha_t^2)/2$
        \STATE $\beta_{t+1} = \alpha_t(1-\alpha_t)/(\alpha_{t+1} + \alpha_t^2)$
        \STATE $\yv^{(t+1)} = \xv^{(t+1)} + \beta_{t+1} (\xv^{(t+1)} - \xv^{(t)})$
\ENDWHILE
\end{algorithmic}

A residual curve as a function of iteration using Nesterov's acceleration is not monotonic. Restarting Nesterov's acceleration when the residual increases significantly improves the convergence rate \citep{o2015adaptive}.

\begin{algorithmic}
\STATE \underline{\bf APG+restart}
\STATE Minimization of $ q = \xv^T \mathcal{W} \xv - \becv^T \xv$
\STATE Initialization: $T= I - \mathcal{W}/L, \sv = \becv/L, \xv^{(0)}, \yv^{(0)}=\xv^{(0)}, \alpha_0 = 0.9$ 
\WHILE {Condition} 
        \STATE $\xv^{(t+1)} =  \mathcal{P}[T \yv^{(t)} + \sv]$
        \STATE $ q^{(t+1)} = (\xv^{(t+1)})^T \mathcal{W} \xv^{(t+1)} - \becv^T \xv^{(t+1)}$
        \STATE $\alpha_{t+1}=(\sqrt{\alpha_t^4 + 4\alpha_t^2} - \alpha_t^2)/2$
        \STATE $\beta_{t+1} = \alpha_t(1-\alpha_t)/(\alpha_{t+1} + \alpha_t^2)$
        \STATE $\yv^{(t+1)} = \xv^{(t+1)} + \beta_{t+1} (\xv^{(t+1)} - \xv^{(t)})$
        \IF {$q^{(t+1)} > q^{(t)}$}
        \STATE $\xv^{(t+1)} =  \mathcal{P}[T \xv^{(t)} + \sv]$
        \STATE $\yv^{(t+1)}=\xv^{(t+1)},\alpha_{t+1} = \alpha_0$
        \ENDIF
\ENDWHILE
\end{algorithmic}

\begin{figure}[!th]
 \begin{center}
   \includegraphics[width=0.5\linewidth]{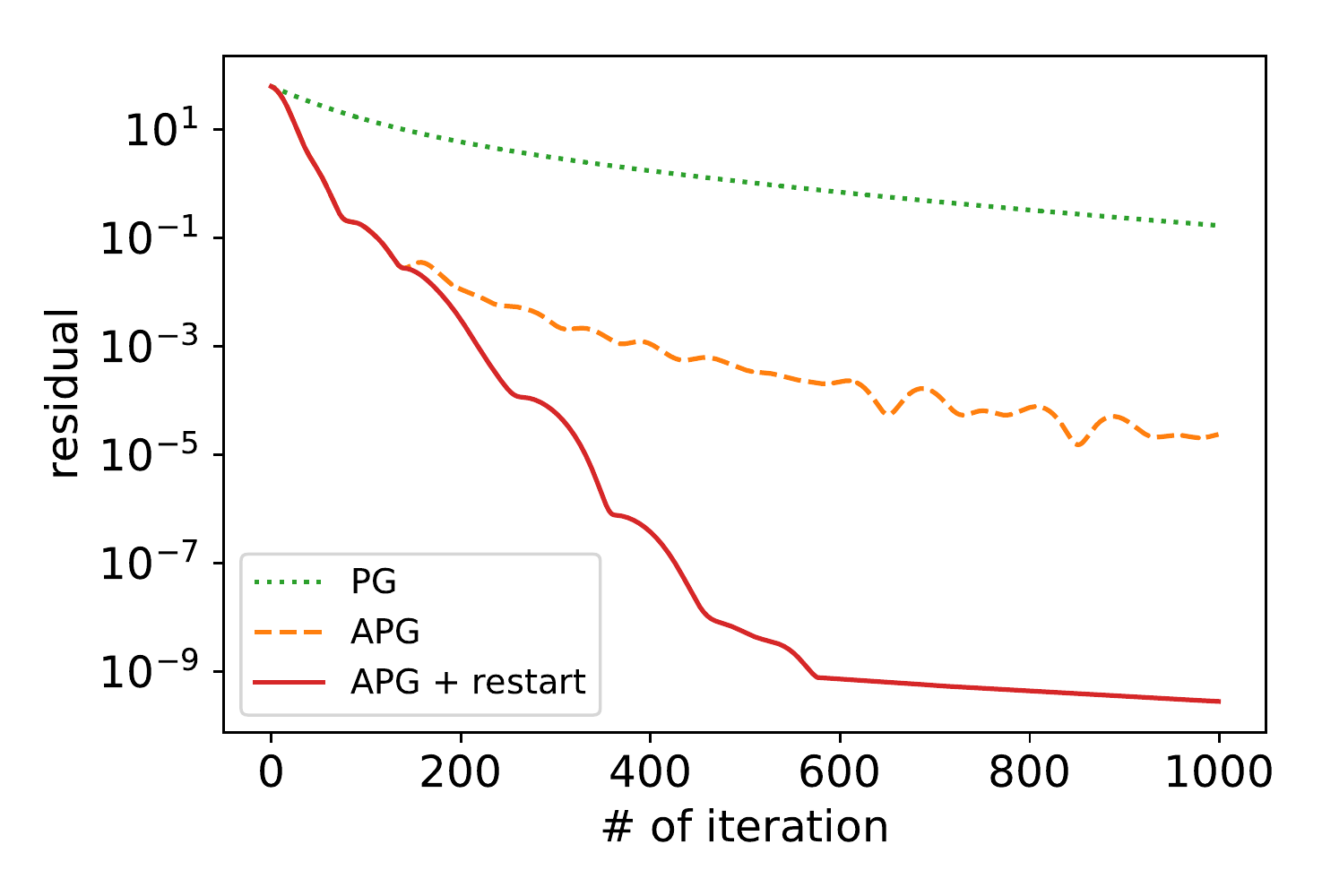}
 \end{center}
 \caption{Residual of various PG solvers as a function of the number of iterations. \label{fig:NNLS}}
\end{figure}

Figure \ref{fig:NNLS} shows a comparison of the above three algorithms for a randomly generated matrix $A$ and a vector $\pv$ as a quadratic problem ($\xv^T \mathcal{W}  \xv - \becv^T \xv$) for $ \mathcal{W} = A^T A$ (100 $\times$ 100 matrix) and $\becv = A^T \pv$. The residual after the $t$-th iteration is defined by $||A \xv^{(t)} - \becv||_2^2$, where $\xv^{(t)}$ is the estimated value after $t$ iterations.

The drawback of the PG and APG method is that convergence is sensitive to the initial point. When all of the components of $ \mathcal{P}[\xv^{0} - \eta \Delta Q] $ {\color{red}are} zero, the algorithm fails. 

\section{Optimization of Weighted NMF by Multiplicative Update}\label{ap:multi}

The multiplicative iterative algorithm \citep{lee2001algorithms} is often used to minimize the cost function of the standard NMF, given in equation (\ref{eq:NMF}). It can be directly derived from the cost function and the Karush--Kuhn--Tucker first-order optimal conditions \citep{cichocki2009nonnegative}. We need to extend the standard multiplicative iterative algorithm to include the geometric kernel $W$ in equation (\ref{eq:WAX}).

First, let us explain the algorithm for the weighted NMF with no regularization term ($R(A) = 0$). Following the derivation of the multiplicative iterative algorithm, we compute the derivative of the cost function of Equation (\ref{eq:WAX}) as 
\begin{eqnarray}
\nabla_A Q &=& W^T W A X X^T - W^T D X^T \\
\nabla_X Q &=& A^T W^T W A X - A^T W^T D.
\end{eqnarray}

To ensure the nonnegativity, we divide the derivative of the cost function into the positive terms and negative terms
\begin{eqnarray}
\nabla Q &=& [\nabla Q]_{+} - [\nabla Q]_{-} = 0,
\end{eqnarray}
where $ [\nabla Q]_{-} \ge 0, [\nabla Q]_{+} \ge 0 $. The multiplicative update is an operation that multiplies $[\nabla Q]_{-}/[\nabla Q]_{+}$ by $A$ or $X$. This procedure can be interpreted as the steepest gradient descent 
\begin{eqnarray}
A &\leftarrow A& - \eta_A \odot \nabla_A Q\\
\eta_A &=& A \oslash [\nabla_A Q]_+
\end{eqnarray}
and
\begin{eqnarray}
X &\leftarrow X& - \eta_X \odot \nabla_X Q \\
\eta_X &=& X \oslash [\nabla_X Q]_+
\end{eqnarray}
where $\odot$ indicates the Hadamard product (the element-wise product of two matrices), and $\oslash$ is the element-wise division. 

The multiplicative iterative algorithm for the weighted NMF with no regularization is given by
\begin{eqnarray}
A_{jk} \leftarrow A_{jk} \frac{[W^T D X^T]_{jk} + \epsilon}{[W^T W A X X^T]_{jk} +\epsilon} \\
X_{kl} \leftarrow X_{kl} \frac{[A^T W^T D]_{kl} + \epsilon}{[A^T W^T W A X]_{kl}+\epsilon},
\end{eqnarray}
where $\epsilon$ is a small value to prevent division by zero. To include the regularization, the derivative of $R(A,X)$ by $A$ or $X$ is needed. For the dual-L2-type, we obtain 
\begin{eqnarray}
\label{eq:derterm}
\nabla_A R(A,X) &=& \lambda_A \, A \\
\nabla_X R(A,X) &=& \lambda_X \, X.
\end{eqnarray}
Because these values remain positive when we take positive values for the initial state, the multiplicative update for the dual L2-type regularization is expressed as 
\begin{eqnarray}
\hbox{U(A):} \,\, A_{jk} &\leftarrow& A_{jk} \frac{[W^T D X^T]_{jk} + \epsilon}{[W^T W A X X^T + \lambda_A A]_{jk} +\epsilon}  \\
\hbox{U(X):} \,\, X_{kl} &\leftarrow& X_{kl} \frac{[A^T W^T D]_{kl} + \epsilon}{[A^T W^T W A X + \lambda_X X]_{kl}+\epsilon}.
\end{eqnarray}

\section{On the Additional Constraints}
In remote sensing, an additional constraint is sometimes applied. For instance, the normalization for a spectrum is expressed by 
\begin{eqnarray}
\label{eq:add_constraints}
\sum_l X_{kl} = 1.
\end{eqnarray}
We found that the constraint of Equation (\ref{eq:add_constraints}) functions as a form of regularization if we combine the constraint with the volume-regularization term. When we use the constraint of Equation (\ref{eq:add_constraints}) with the volume-regularization term, the effect of the volume-regularization vanishes. We do not recommend the use of the constraint of Equation (\ref{eq:add_constraints}) is used in our case. 

\bibliography{sample63}{}

\begin{thebibliography}{}
\expandafter\ifx\csname natexlab\endcsname\relax\def\natexlab#1{#1}\fi
\providecommand{\url}[1]{\href{#1}{#1}}
\providecommand{\dodoi}[1]{doi:~\href{http://doi.org/#1}{\nolinkurl{#1}}}
\providecommand{\doeprint}[1]{\href{http://ascl.net/#1}{\nolinkurl{http://ascl.net/#1}}}
\providecommand{\doarXiv}[1]{\href{https://arxiv.org/abs/#1}{\nolinkurl{https://arxiv.org/abs/#1}}}

\bibitem[{{Aizawa} {et~al.}(2020){Aizawa}, {Kawahara}, \& {Fan}}]{Aizawa}
{Aizawa}, M., {Kawahara}, H., \& {Fan}, S. 2020, under review

\bibitem[{Ang \& Gillis(2018)}]{ang2018volume}
Ang, M.~A., \& Gillis, N. 2018, in 2018 9th Workshop on Hyperspectral Image and
  Signal Processing: Evolution in Remote Sensing (WHISPERS), IEEE, 1--5

\bibitem[{{Ang} \& {Gillis}(2019)}]{2019arXiv190304362A}
{Ang}, M.~S., \& {Gillis}, N. 2019, arXiv e-prints, arXiv:1903.04362.
\newblock \doarXiv{1903.04362}

\bibitem[{Baldridge {et~al.}(2009)Baldridge, Hook, Grove, \&
  Rivera}]{baldridge2009aster}
Baldridge, A.~M., Hook, S., Grove, C., \& Rivera, G. 2009, Remote Sensing of
  Environment, 113, 711

\bibitem[{{Berdyugina} \& {Kuhn}(2019)}]{2019AJ....158..246B}
{Berdyugina}, S.~V., \& {Kuhn}, J.~R. 2019, \aj, 158, 246,
  \dodoi{10.3847/1538-3881/ab2df3}

\bibitem[{Cichocki {et~al.}(2009)Cichocki, Zdunek, Phan, \&
  Amari}]{cichocki2009nonnegative}
Cichocki, A., Zdunek, R., Phan, A.~H., \& Amari, S.-i. 2009, Nonnegative matrix
  and tensor factorizations: applications to exploratory multi-way data
  analysis and blind source separation (John Wiley \& Sons)

\bibitem[{{Cowan} {et~al.}(2013){Cowan}, {Fuentes}, \&
  {Haggard}}]{2013MNRAS.434.2465C}
{Cowan}, N.~B., {Fuentes}, P.~A., \& {Haggard}, H.~M. 2013, \mnras, 434, 2465,
  \dodoi{10.1093/mnras/stt1191}

\bibitem[{{Cowan} \& {Strait}(2013)}]{2013ApJ...765L..17C}
{Cowan}, N.~B., \& {Strait}, T.~E. 2013, \apjl, 765, L17,
  \dodoi{10.1088/2041-8205/765/1/L17}

\bibitem[{{Cowan} {et~al.}(2009){Cowan}, {Agol}, {Meadows}, {Robinson},
  {Livengood}, {Deming}, {Lisse}, {A'Hearn}, {Wellnitz}, {Seager},
  {Charbonneau}, \& {EPOXI Team}}]{2009ApJ...700..915C}
{Cowan}, N.~B., {Agol}, E., {Meadows}, V.~S., {et~al.} 2009, \apj, 700, 915,
  \dodoi{10.1088/0004-637X/700/2/915}

\bibitem[{{Craig}(1994)}]{1994ITGRS..32..542C}
{Craig}, M.~D. 1994, IEEE Transactions on Geoscience and Remote Sensing, 32,
  542, \dodoi{10.1109/36.297973}

\bibitem[{{Fan} {et~al.}(2019){Fan}, {Li}, {Li}, {Bartlett}, {Jiang}, {Natraj},
  {Crisp}, \& {Yung}}]{2019ApJ...882L...1F}
{Fan}, S., {Li}, C., {Li}, J.-Z., {et~al.} 2019, \apjl, 882, L1,
  \dodoi{10.3847/2041-8213/ab3a49}

\bibitem[{{Farr} {et~al.}(2018){Farr}, {Farr}, {Cowan}, {Haggard}, \&
  {Robinson}}]{2018AJ....156..146F}
{Farr}, B., {Farr}, W.~M., {Cowan}, N.~B., {Haggard}, H.~M., \& {Robinson}, T.
  2018, \aj, 156, 146, \dodoi{10.3847/1538-3881/aad775}

\bibitem[{{Ford} {et~al.}(2001){Ford}, {Seager}, \&
  {Turner}}]{2001Natur.412..885F}
{Ford}, E.~B., {Seager}, S., \& {Turner}, E.~L. 2001, \nat, 412, 885,
  \dodoi{10.1038/35091009}

\bibitem[{{Fu} {et~al.}(2019){Fu}, {Huang}, {Sidiropoulos}, \&
  {Ma}}]{2019ISPM...36...59F}
{Fu}, X., {Huang}, K., {Sidiropoulos}, N.~D., \& {Ma}, W.-K. 2019, IEEE Signal
  Processing Magazine, 36, 59, \dodoi{10.1109/MSP.2018.2877582}

\bibitem[{{Fu} {et~al.}(2015){Fu}, {Ma}, {Huang}, \&
  {Sidiropoulos}}]{2015ITSP...63.2306F}
{Fu}, X., {Ma}, W.-K., {Huang}, K., \& {Sidiropoulos}, N.~D. 2015, IEEE
  Transactions on Signal Processing, 63, 2306, \dodoi{10.1109/TSP.2015.2404577}

\bibitem[{{Fujii} \& {Kawahara}(2012)}]{2012ApJ...755..101F}
{Fujii}, Y., \& {Kawahara}, H. 2012, \apj, 755, 101,
  \dodoi{10.1088/0004-637X/755/2/101}

\bibitem[{{Fujii} {et~al.}(2011){Fujii}, {Kawahara}, {Suto}, {Fukuda},
  {Nakajima}, {Livengood}, \& {Turner}}]{2011ApJ...738..184F}
{Fujii}, Y., {Kawahara}, H., {Suto}, Y., {et~al.} 2011, \apj, 738, 184,
  \dodoi{10.1088/0004-637X/738/2/184}

\bibitem[{{Fujii} {et~al.}(2010){Fujii}, {Kawahara}, {Suto}, {Taruya},
  {Fukuda}, {Nakajima}, \& {Turner}}]{2010ApJ...715..866F}
---. 2010, \apj, 715, 866, \dodoi{10.1088/0004-637X/715/2/866}

\bibitem[{{Fujii} {et~al.}(2017){Fujii}, {Lustig-Yaeger}, \&
  {Cowan}}]{2017AJ....154..189F}
{Fujii}, Y., {Lustig-Yaeger}, J., \& {Cowan}, N.~B. 2017, \aj, 154, 189,
  \dodoi{10.3847/1538-3881/aa89f1}

\bibitem[{{G{\'o}rski} {et~al.}(2005){G{\'o}rski}, {Hivon}, {Banday},
  {Wandelt}, {Hansen}, {Reinecke}, \& {Bartelmann}}]{2005ApJ...622..759G}
{G{\'o}rski}, K.~M., {Hivon}, E., {Banday}, A.~J., {et~al.} 2005, \apj, 622,
  759, \dodoi{10.1086/427976}

\bibitem[{{Haggard} \& {Cowan}(2018)}]{2018MNRAS.478..371H}
{Haggard}, H.~M., \& {Cowan}, N.~B. 2018, \mnras, 478, 371,
  \dodoi{10.1093/mnras/sty1019}

\bibitem[{{Jiang} {et~al.}(2018){Jiang}, {Zhai}, {Herman}, {Zhai}, {Hu}, {Su},
  {Natraj}, {Li}, {Xu}, \& {Yung}}]{2018AJ....156...26J}
{Jiang}, J.~H., {Zhai}, A.~J., {Herman}, J., {et~al.} 2018, \aj, 156, 26,
  \dodoi{10.3847/1538-3881/aac6e2}

\bibitem[{{Kawahara}(2016)}]{2016ApJ...822..112K}
{Kawahara}, H. 2016, \apj, 822, 112, \dodoi{10.3847/0004-637X/822/2/112}

\bibitem[{{Kawahara} \& {Fujii}(2010)}]{2010ApJ...720.1333K}
{Kawahara}, H., \& {Fujii}, Y. 2010, \apj, 720, 1333,
  \dodoi{10.1088/0004-637X/720/2/1333}

\bibitem[{{Kawahara} \& {Fujii}(2011)}]{2011ApJ...739L..62K}
---. 2011, \apjl, 739, L62, \dodoi{10.1088/2041-8205/739/2/L62}

\bibitem[{Kim {et~al.}(2014)Kim, He, \& Park}]{kim2014algorithms}
Kim, J., He, Y., \& Park, H. 2014, Journal of Global Optimization, 58, 285

\bibitem[{Lawson \& Hanson(1995)}]{lawson1995solving}
Lawson, C.~L., \& Hanson, R.~J. 1995, Solving least squares problems, Vol.~15
  (Siam)

\bibitem[{Lee \& Seung(2001)}]{lee2001algorithms}
Lee, D.~D., \& Seung, H.~S. 2001, in Advances in neural information processing
  systems, 556--562

\bibitem[{{Lin} {et~al.}(2015){Lin}, {Ma}, {Li}, {Chi}, \&
  {Ambikapathi}}]{2015ITGRS..53.5530L}
{Lin}, C.-H., {Ma}, W.-K., {Li}, W.-C., {Chi}, C.-Y., \& {Ambikapathi}, A.
  2015, IEEE Transactions on Geoscience and Remote Sensing, 53, 5530,
  \dodoi{10.1109/TGRS.2015.2424719}

\bibitem[{{Luger} {et~al.}(2019){Luger}, {Bedell}, {Vanderspek}, \&
  {Burke}}]{2019arXiv190312182L}
{Luger}, R., {Bedell}, M., {Vanderspek}, R., \& {Burke}, C.~J. 2019, arXiv
  e-prints, arXiv:1903.12182.
\newblock \doarXiv{1903.12182}

\bibitem[{{Lustig-Yaeger} {et~al.}(2018){Lustig-Yaeger}, {Meadows}, {Tovar
  Mendoza}, {Schwieterman}, {Fujii}, {Luger}, \&
  {Robinson}}]{2018AJ....156..301L}
{Lustig-Yaeger}, J., {Meadows}, V.~S., {Tovar Mendoza}, G., {et~al.} 2018, \aj,
  156, 301, \dodoi{10.3847/1538-3881/aaed3a}

\bibitem[{{McLinden} {et~al.}(1997){McLinden}, {McConnell}, {Griffioen},
  {McElroy}, \& {Pfister}}]{1997JGR...10218801M}
{McLinden}, C.~A., {McConnell}, J.~C., {Griffioen}, E., {McElroy}, C.~T., \&
  {Pfister}, L. 1997, \jgr, 102, 18,801, \dodoi{10.1029/97JD01079}

\bibitem[{{Nakagawa} {et~al.}(2020){Nakagawa}, {Kodama}, \&
  {Ishiwatari}}]{Nakagawa}
{Nakagawa}, Y., {Kodama}, T., \& {Ishiwatari}, M. 2020, under review

\bibitem[{Nesterov(1983)}]{nesterov1983method}
Nesterov, Y.~E. 1983, in Dokl. akad. nauk Sssr, Vol. 269, 543--547

\bibitem[{{Oakley} \& {Cash}(2009)}]{2009ApJ...700.1428O}
{Oakley}, P.~H.~H., \& {Cash}, W. 2009, \apj, 700, 1428,
  \dodoi{10.1088/0004-637X/700/2/1428}

\bibitem[{O’donoghue \& Candes(2015)}]{o2015adaptive}
O’donoghue, B., \& Candes, E. 2015, Foundations of computational mathematics,
  15, 715

\bibitem[{Paatero \& Tapper(1994)}]{paatero1994positive}
Paatero, P., \& Tapper, U. 1994, Environmetrics, 5, 111

\bibitem[{{Pall{\'e}} {et~al.}(2008){Pall{\'e}}, {Ford}, {Seager},
  {Monta{\~n}{\'e}s-Rodr{\'{\i}}guez}, \& {Vazquez}}]{2008ApJ...676.1319P}
{Pall{\'e}}, E., {Ford}, E.~B., {Seager}, S.,
  {Monta{\~n}{\'e}s-Rodr{\'{\i}}guez}, P., \& {Vazquez}, M. 2008, \apj, 676,
  1319, \dodoi{10.1086/528677}

\bibitem[{Schachtner {et~al.}(2009)Schachtner, P{\"o}ppel, Tom{\'e}, \&
  Lang}]{schachtner2009minimum}
Schachtner, R., P{\"o}ppel, G., Tom{\'e}, A.~M., \& Lang, E.~W. 2009, in
  International Conference on Independent Component Analysis and Signal
  Separation, Springer, 106--113

\bibitem[{{Schwartz} {et~al.}(2016){Schwartz}, {Sekowski}, {Haggard},
  {Pall{\'e}}, \& {Cowan}}]{2016MNRAS.457..926S}
{Schwartz}, J.~C., {Sekowski}, C., {Haggard}, H.~M., {Pall{\'e}}, E., \&
  {Cowan}, N.~B. 2016, \mnras, 457, 926, \dodoi{10.1093/mnras/stw068}

\bibitem[{Vavasis(2009)}]{vavasis2009complexity}
Vavasis, S.~A. 2009, SIAM Journal on Optimization, 20, 1364

\bibitem[{Xiang {et~al.}(2015)Xiang, Peng, \& Yang}]{xiang2015blind}
Xiang, Y., Peng, D., \& Yang, Z. 2015, Blind source separation: dependent
  component analysis (Springer)

\bibitem[{Zhou {et~al.}(2011)Zhou, Xie, Yang, Yang, \& He}]{zhou2011minimum}
Zhou, G., Xie, S., Yang, Z., Yang, J.-M., \& He, Z. 2011, IEEE transactions on
  neural networks, 22, 1626

\end{thebibliography}
\bibliographystyle{aasjournal}
\end{document}